\documentclass[twocolumn]{aastex63}
\submitjournal{The Astrophysical Journal}
\usepackage{natbib,graphicx,lineno}
\usepackage{hyperref}
\usepackage[utf8]{inputenc}  
\hypersetup{
    colorlinks=true,
    linkcolor=blue,
    filecolor=magenta,      
    urlcolor=blue,
    citecolor=brown
}

\usepackage{booktabs}
\usepackage{lineno}
\usepackage{xspace}   

\newcommand{\Flat}{\emph{Fermi}-LAT\xspace}
\newcommand{\fhit}{\emph{fHit}\xspace}

\shorttitle{Nova V392 Per with \emph{Fermi} and HAWC}
\shortauthors{HAWC Collaboration}

\begin{document}
\title{$\gamma$-ray Emission from Classical Nova V392 Per: Measurements from \emph{Fermi} and HAWC}
\author[0000-0003-0197-5646]{A.~Albert}
\affiliation{Physics Division, Los Alamos National Laboratory, Los Alamos, NM, USA }
\author[0000-0001-8749-1647]{R.~Alfaro}
\affiliation{Instituto de F'{i}sica, Universidad Nacional Autónoma de México, Ciudad de Mexico, Mexico }
\author{C.~Alvarez}
\affiliation{Universidad Autónoma de Chiapas, Tuxtla Gutiérrez, Chiapas, México}
\author{J.C.~Arteaga-Velázquez}
\affiliation{Universidad Michoacana de San Nicolás de Hidalgo, Morelia, Mexico }
\author[0000-0002-4020-4142]{D.~Avila Rojas}
\affiliation{Instituto de F'{i}sica, Universidad Nacional Autónoma de México, Ciudad de Mexico, Mexico }
\author[0000-0002-2084-5049]{H.A.~Ayala Solares}
\affiliation{Department of Physics, Pennsylvania State University, University Park, PA, USA }
\author[0000-0002-5529-6780]{R.~Babu}
\affiliation{Department of Physics, Michigan Technological University, Houghton, MI, USA }
\author[0000-0003-3207-105X]{E.~Belmont-Moreno}
\affiliation{Instituto de F'{i}sica, Universidad Nacional Autónoma de México, Ciudad de Mexico, Mexico }
\author{C. Blochwitz}
\affiliation{Department of Physics and Astronomy, Michigan State University, East Lansing, MI, USA }
\affiliation{Center for Data Intensive and Time Domain Astronomy, Department of Physics and Astronomy, Michigan State University, East Lansing MI 48824, USA}
\author[0000-0002-4042-3855]{K.S.~Caballero-Mora}
\affiliation{Universidad Autónoma de Chiapas, Tuxtla Gutiérrez, Chiapas, México}
\author[0000-0003-2158-2292]{T.~Capistrán}
\affiliation{Instituto de Astronom'{i}a, Universidad Nacional Autónoma de México, Ciudad de Mexico, Mexico }
\author[0000-0002-8553-3302]{A.~Carramiñana}
\affiliation{Instituto Nacional de Astrof'{i}sica, Óptica y Electrónica, Puebla, Mexico }
\author[0000-0002-6144-9122]{S.~Casanova}
\affiliation{Instytut Fizyki Jadrowej im Henryka Niewodniczanskiego Polskiej Akademii Nauk, IFJ-PAN, Krakow, Poland }
\author{O.~Chaparro-Amaro}
\affiliation{Centro de Investigaci'on en Computaci'on, Instituto Polit'ecnico Nacional, M'exico City, M'exico.}
\author[0000-0002-7607-9582]{U.~Cotti}
\affiliation{Universidad Michoacana de San Nicolás de Hidalgo, Morelia, Mexico }
\author[0000-0002-1132-871X]{J.~Cotzomi}
\affiliation{Facultad de Ciencias F'{i}sico Matemáticas, Benemérita Universidad Autónoma de Puebla, Puebla, Mexico }
\author[0000-0001-9643-4134]{E.~De la Fuente}
\affiliation{Departamento de F'{i}sica, Centro Universitario de Ciencias Exactase Ingenierias, Universidad de Guadalajara, Guadalajara, Mexico }
\author[0000-0002-8528-9573]{C.~de León}
\affiliation{Universidad Michoacana de San Nicolás de Hidalgo, Morelia, Mexico }
\author[0000-0002-7747-754X]{S.~Coutiño de León}
\affiliation{Department of Physics, University of Wisconsin-Madison, Madison, WI, USA }
\author{R.~Diaz Hernandez}
\affiliation{Instituto Nacional de Astrof'{i}sica, Óptica y Electrónica, Puebla, Mexico }
\author[0000-0001-8451-7450]{B.L.~Dingus}
\affiliation{Physics Division, Los Alamos National Laboratory, Los Alamos, NM, USA }
\author[0000-0002-2987-9691]{M.A.~DuVernois}
\affiliation{Department of Physics, University of Wisconsin-Madison, Madison, WI, USA }
\author[0000-0003-2169-0306]{M.~Durocher}
\affiliation{Physics Division, Los Alamos National Laboratory, Los Alamos, NM, USA }
\author{J.C.~Díaz-Vélez}
\affiliation{Departamento de F'{i}sica, Centro Universitario de Ciencias Exactase Ingenierias, Universidad de Guadalajara, Guadalajara, Mexico }
\author[0000-0001-5737-1820]{K.~Engel}
\affiliation{Department of Physics, University of Maryland, College Park, MD, USA }
\author[0000-0001-7074-1726]{C.~Espinoza}
\affiliation{Instituto de F'{i}sica, Universidad Nacional Autónoma de México, Ciudad de Mexico, Mexico }
\author[0000-0002-8246-4751]{K.L.~Fan}
\affiliation{Department of Physics, University of Maryland, College Park, MD, USA }
\author[0000-0002-5387-8138]{K.~Fang}
\affiliation{Department of Physics, University of Wisconsin-Madison, Madison, WI, USA }
\author[0000-0002-0173-6453]{N.~Fraija}
\affiliation{Instituto de Astronom'{i}a, Universidad Nacional Autónoma de México, Ciudad de Mexico, Mexico }
\author[0000-0002-4188-5584]{J.A.~García-González}
\affiliation{Tecnologico de Monterrey, Escuela de Ingenier\'{i}a y Ciencias, Ave. Eugenio Garza Sada 2501, Monterrey, N.L., Mexico, 64849}
\author[0000-0003-1122-4168]{F.~Garfias}
\affiliation{Instituto de Astronom'{i}a, Universidad Nacional Autónoma de México, Ciudad de Mexico, Mexico }
\author[0000-0002-5209-5641]{M.M.~González}
\affiliation{Instituto de Astronom'{i}a, Universidad Nacional Autónoma de México, Ciudad de Mexico, Mexico }
\author[0000-0002-9790-1299]{J.A.~Goodman}
\affiliation{Department of Physics, University of Maryland, College Park, MD, USA }
\author[0000-0001-9844-2648]{J.P.~Harding}
\affiliation{Physics Division, Los Alamos National Laboratory, Los Alamos, NM, USA }
\author[0000-0002-2565-8365]{S.~Hernandez}
\affiliation{Instituto de F'{i}sica, Universidad Nacional Autónoma de México, Ciudad de Mexico, Mexico }
\author[0000-0002-1031-7760]{J.~Hinton}
\affiliation{Max-Planck Institute for Nuclear Physics, 69117 Heidelberg, Germany}
\author[0000-0002-5447-1786]{D.~Huang}
\affiliation{Department of Physics, Michigan Technological University, Houghton, MI, USA }
\author[0000-0002-5527-7141]{F.~Hueyotl-Zahuantitla}
\affiliation{Universidad Autónoma de Chiapas, Tuxtla Gutiérrez, Chiapas, México}
\author{P.~Hüntemeyer}
\affiliation{Department of Physics, Michigan Technological University, Houghton, MI, USA }
\author[0000-0001-5811-5167]{A.~Iriarte}
\affiliation{Instituto de Astronom'{i}a, Universidad Nacional Autónoma de México, Ciudad de Mexico, Mexico }
\author[0000-0003-4467-3621]{V.~Joshi}
\affiliation{Erlangen Centre for Astroparticle Physics, Friedrich-Alexander-Universit\"at Erlangen-N\"urnberg, Erlangen, Germany}
\author[0000-0001-6336-5291]{A.~Lara}
\affiliation{Instituto de Geof'{i}sica, Universidad Nacional Autónoma de México, Ciudad de Mexico, Mexico }
\author[0000-0002-2467-5673]{W.H.~Lee}
\affiliation{Instituto de Astronom'{i}a, Universidad Nacional Autónoma de México, Ciudad de Mexico, Mexico }
\author[0000-0003-2696-947X]{J.T.~Linnemann}
\affiliation{Department of Physics and Astronomy, Michigan State University, East Lansing, MI, USA }
\author[0000-0001-8825-3624]{A.L.~Longinotti}
\affiliation{Instituto de Astronom'{i}a, Universidad Nacional Autónoma de México, Ciudad de Mexico, Mexico }
\author[0000-0003-2810-4867]{G.~Luis-Raya}
\affiliation{Universidad Politecnica de Pachuca, Pachuca, Hgo, Mexico }
\author[0000-0003-3751-5617]{J.~Lundeen}
\affiliation{Department of Physics and Astronomy, Michigan State University, East Lansing, MI, USA }
\author[0000-0001-8088-400X]{K.~Malone}
\affiliation{Physics Division, Los Alamos National Laboratory, Los Alamos, NM, USA }
\author[0000-0001-9077-4058]{V.~Marandon}
\affiliation{Max-Planck Institute for Nuclear Physics, 69117 Heidelberg, Germany}
\author[0000-0001-9052-856X]{O.~Martinez}
\affiliation{Facultad de Ciencias F'{i}sico Matemáticas, Benemérita Universidad Autónoma de Puebla, Puebla, Mexico }
\author[0000-0002-2824-3544]{J.~Martínez-Castro}
\affiliation{Centro de Investigaci'on en Computaci'on, Instituto Polit'ecnico Nacional, M'exico City, M'exico.}
\author[0000-0002-2610-863X]{J.A.~Matthews}
\affiliation{Dept of Physics and Astronomy, University of New Mexico, Albuquerque, NM, USA }
\author[0000-0002-8390-9011]{P.~Miranda-Romagnoli}
\affiliation{Universidad Autónoma del Estado de Hidalgo, Pachuca, Mexico }
\author[0000-0001-9361-0147]{J.A.~Morales-Soto}
\affiliation{Universidad Michoacana de San Nicolás de Hidalgo, Morelia, Mexico }
\author[0000-0002-1114-2640]{E.~Moreno}
\affiliation{Facultad de Ciencias F'{i}sico Matemáticas, Benemérita Universidad Autónoma de Puebla, Puebla, Mexico }
\author[0000-0002-7675-4656]{M.~Mostafá}
\affiliation{Department of Physics, Pennsylvania State University, University Park, PA, USA }
\author[0000-0003-0587-4324]{A.~Nayerhoda}
\affiliation{Instytut Fizyki Jadrowej im Henryka Niewodniczanskiego Polskiej Akademii Nauk, IFJ-PAN, Krakow, Poland }
\author[0000-0003-1059-8731]{L.~Nellen}
\affiliation{Instituto de Ciencias Nucleares, Universidad Nacional Autónoma de Mexico, Ciudad de Mexico, Mexico }
\author[0000-0001-9428-7572]{M.~Newbold}
\affiliation{Department of Physics and Astronomy, University of Utah, Salt Lake City, UT, USA }
\author[0000-0002-6859-3944]{M.U.~Nisa}
\affiliation{Department of Physics and Astronomy, Michigan State University, East Lansing, MI, USA }
\author[0000-0001-7099-108X]{R.~Noriega-Papaqui}
\affiliation{Universidad Autónoma del Estado de Hidalgo, Pachuca, Mexico }
\author[0000-0002-5448-7577]{N.~Omodei}
\affiliation{Department of Physics, Stanford University: Stanford, CA 94305–4060, USA}
\author{A.~Peisker}
\affiliation{Department of Physics and Astronomy, Michigan State University, East Lansing, MI, USA }
\author[0000-0002-8774-8147]{Y.~Pérez Araujo}
\affiliation{Instituto de Astronom'{i}a, Universidad Nacional Autónoma de México, Ciudad de Mexico, Mexico }
\author[0000-0001-5998-4938]{E.G.~Pérez-Pérez}
\affiliation{Universidad Politecnica de Pachuca, Pachuca, Hgo, Mexico }
\author[0000-0002-6524-9769]{C.D.~Rho}
\affiliation{University of Seoul, Seoul, Rep. of Korea}
\author[0000-0003-1327-0838]{D.~Rosa-González}
\affiliation{Instituto Nacional de Astrof'{i}sica, Óptica y Electrónica, Puebla, Mexico }
\author[0000-0001-6939-7825]{E.~Ruiz-Velasco}
\affiliation{Max-Planck Institute for Nuclear Physics, 69117 Heidelberg, Germany}
\author[0000-0002-9312-9684]{D.~Salazar-Gallegos}
\affiliation{Department of Physics and Astronomy, Michigan State University, East Lansing, MI, USA }
\author[0000-0002-8610-8703]{F.~Salesa Greus}
\affiliation{Instytut Fizyki Jadrowej im Henryka Niewodniczanskiego Polskiej Akademii Nauk, IFJ-PAN, Krakow, Poland }
\affiliation{Instituto de Física Corpuscular, CSIC, Universitat de València, E-46980, Paterna, Valencia, Spain}
\author[0000-0001-6079-2722]{A.~Sandoval}
\affiliation{Instituto de F'{i}sica, Universidad Nacional Autónoma de México, Ciudad de Mexico, Mexico }
\author{J.~Serna-Franco}
\affiliation{Instituto de F'{i}sica, Universidad Nacional Autónoma de México, Ciudad de Mexico, Mexico }
\author[0000-0002-1012-0431]{A.J.~Smith}
\affiliation{Department of Physics, University of Maryland, College Park, MD, USA }
\author{Y.~Son}
\affiliation{University of Seoul, Seoul, Rep. of Korea}
\author[0000-0002-1492-0380]{R.W.~Springer}
\affiliation{Department of Physics and Astronomy, University of Utah, Salt Lake City, UT, USA }
\author{
{O.~Tibolla}}
\affiliation{Universidad Politecnica de Pachuca, Pachuca, Hgo, Mexico }
\author[0000-0001-9725-1479]{K.~Tollefson}
\affiliation{Department of Physics and Astronomy, Michigan State University, East Lansing, MI, USA }
\author[0000-0002-1689-3945]{I.~Torres}
\affiliation{Instituto Nacional de Astrof'{i}sica, Óptica y Electrónica, Puebla, Mexico }
\author{R.~Torres-Escobedo}
\affiliation{Tsung-Dao Lee Institute and School of Physics and Astronomy, Shanghai Jiao Tong University, Shanghai, China}
\author[0000-0003-1068-6707]{R.~Turner}
\affiliation{Department of Physics, Michigan Technological University, Houghton, MI, USA }
\author[0000-0002-2748-2527]{F.~Ureña-Mena}
\affiliation{Instituto Nacional de Astrof'{i}sica, Óptica y Electrónica, Puebla, Mexico }
\author[0000-0001-6876-2800]{L.~Villaseñor}
\affiliation{Facultad de Ciencias F'{i}sico Matemáticas, Benemérita Universidad Autónoma de Puebla, Puebla, Mexico }
\author[0000-0001-6798-353X]{X.~Wang}
\affiliation{Department of Physics, Michigan Technological University, Houghton, MI, USA }
\author[0000-0002-6623-0277]{E.~Willox}
\affiliation{Department of Physics, University of Maryland, College Park, MD, USA }
\author[0000-0001-9976-2387]{A.~Zepeda}
\affiliation{Physics Department, Centro de Investigacion y de Estudios Avanzados del IPN, Mexico City, Mexico }
\author[0000-0003-0513-3841]{H.~Zhou}
\affiliation{Tsung-Dao Lee Institute and School of Physics and Astronomy, Shanghai Jiao Tong University, Shanghai, China}

\collaboration{150}{HAWC Collaboration}

\affiliation{Center for Data Intensive and Time Domain Astronomy, Department of Physics and Astronomy, Michigan State University, East Lansing MI 48824, USA}
\author[0000-0002-8400-3705]{L.~Chomiuk} 
\affiliation{Center for Data Intensive and Time Domain Astronomy, Department of Physics and Astronomy, Michigan State University, East Lansing MI 48824, USA}
\author[0000-0001-8525-3442]{E.~Aydi}
\affiliation{Center for Data Intensive and Time Domain Astronomy, Department of Physics and Astronomy, Michigan State University, East Lansing MI 48824, USA}
\author[0000-0001-8229-2024]{K.L.~Li}
\affiliation{Institute of Astronomy, National Cheng Kung University, Tainan 70101, Taiwan}
\author[0000-0002-4670-7509]{B.D.~Metzger}
\affiliation{Department of Physics and Columbia Astrophysics Laboratory, Columbia University, Pupin Hall, New York, NY 10027, USA}
\affiliation{Center for Computational Astrophysics, Flatiron Institute, 162 5th Ave, New York, NY 10010, USA} 
\author[0000-0003-1336-4746]{I.~Vurm}
\affil{Tartu Observatory, Tartu University, 61602 T{$\bar o$}ravere, Tartumaa, Estonia}
\correspondingauthor{James Linnemann}
\email{linneman@msu.edu}


\begin{abstract}
   
This paper reports on the $\gamma$-ray properties of the 2018 Galactic nova V392 Per, spanning photon energies $\sim$0.1 GeV -- 100 TeV by combining observations from the \emph{Fermi Gamma-ray Space Telescope} and the HAWC Observatory.  In one of the most rapidly evolving $\gamma$-ray signals yet observed for a nova,
GeV $\gamma$ rays with a power law spectrum with index $\Gamma = 2.0 \pm 0.1$
were detected over eight days following V392 Per's optical maximum.
HAWC observations constrain the TeV $\gamma$-ray signal during this time and also before and after.
We observe no statistically significant evidence of TeV $\gamma$-ray emission from V392 Per, but present flux limits. Tests of the extension of the \Flat spectrum to energies above 5 TeV are disfavored by 2 standard deviations (95\%) or more. We fit V392 Per's GeV $\gamma$ rays with hadronic acceleration models, incorporating optical observations, and compare the calculations with HAWC limits.   
\end{abstract}


\section{Introduction} \label{intro}

A classical nova is an explosion in a binary star, occurring on a white dwarf that has accreted mass from a companion star until enough material has accumulated for a thermonuclear runaway. The subsequent eruption ejects the bulk of the accreted material at a few thousand km s$^{-1}$ \citep{Gallagher&Starrfield78, Bode&Evans08, Chomiuk+21araa}. 
{Classical} novae have long been observed at optical wavelengths, but in 2010 the Large Area Telescope (LAT) on the \emph{Fermi Gamma-ray Space Telescope} observed GeV $\gamma$-ray emission from the  nova eruption of V407 Cyg \citep{Abdo_etal10}.  
Although novae had not been expected to produce GeV $\gamma$-ray photons \citep[e.g.,][]{Chomiuk_etal19}, 
 \Flat has since detected $\gamma$ rays in the energy range of .1 to 10 GeV from over a dozen Galactic novae \citep{Ackermann+14, Cheung+16, Franckowiak+18, Gordon+21,  Chomiuk+21araa}. 

These GeV $\gamma$ rays are thought to be the by-product of relativistic particles accelerated by shocks in the nova ejecta \citep{Chomiuk+21araa}. In a few systems with evolved companions, the shocks may mark the interaction of the nova ejecta with pre-existing circumbinary material \citep{Abdo_etal10, Delgado&Hernanz19}, but in novae with 
{main sequence star companions}, the shocks are thought to be internal to the nova ejecta themselves \citep{Chomiuk+14, Martin+18}. The $\gamma$ rays are surprisingly luminous, weighing in at $\sim$0.1--1\% of the bolometric luminosity \citep{Metzger+15}. The implication is that the shocks must be very energetic (rivaling the luminosity of the white dwarf) and/or very efficient at producing $\gamma$ rays. 
 In addition, \citep{Metzger+16} predicts these events could generate photon energies up to 10 TeV, depending on details of the shocks---although TeV emission has yet to be detected from novae. 

This work uses  \Flat to establish the GeV $\gamma$-ray properties of the 2018 nova V392~Per, and then uses archival data from the High-Altitude Water Cherenkov (HAWC) Observatory to see whether this classical nova also produces TeV $\gamma$ rays. 
V392 Per before its 2018 classical nova outburst was known as a 17th (apparent) magnitude dwarf nova discovered in 1970, which had occasional outbursts of up to 3 magnitudes \citep{v392progenitor.18}.  The system has a short 3.2 day period ~\citep{v392.period}.  Although uncommon for dwarf novae, in 2018 V392 Per underwent a classical nova eruption, its brightness rising by 11 magnitudes ($\approx \times 25,000$).   

Two \Flat-detected novae have previously been examined for photon emission in the TeV band using air Cherenkov telescopes.  VERITAS observed V407 Cyg \citep{Aliu_2012} and 
MAGIC observed the nova V339 Del \citep{Ahnen_etal15},  both reporting upper limits on TeV flux. 
Because HAWC is in operation over 95\% of the time, HAWC can search for emission before, during, and after the GeV emission peak for any nova in its field of view.


In \S 2, we discuss the sample of novae we considered and our selection process. In \S 3 we present the GeV properties of the V392 Per nova.  In \S 4, we discuss HAWC analysis techniques and present significance maps of the nova eruption of V392 Per. In \S 5, we consider whether the GeV spectrum continues into the TeV region. \S 6 presents our energy-dependent flux limits.  \S 7 considers systematic uncertainties of the HAWC results. \S 8 describes modeling of V392 Per, and \S 9 presents our conclusions from the study.

\section{ Selection of TeV Nova Candidates for Study with HAWC} 

To study novae most likely to be visible in the TeV band, we  focused on sources that have been detected in the GeV $\gamma$-ray band with \Flat. We considered novae detected with $\gtrsim 3 \sigma$ significance in their time-integrated LAT light curves, as presented in Table S1 of \citep{Chomiuk+21araa}.\footnote{See also \href{https://asd.gsfc.nasa.gov/Koji.Mukai/novae/novae.html}{https://asd.gsfc.nasa.gov/Koji.Mukai/novae/novae.html}} 


The HAWC Observatory is located on the flanks of the Sierra Negra volcano in the state of Puebla, Mexico at an altitude of 4100 m. HAWC has 300 water tanks, each of which contain 4 photomultiplier tubes (PMTs) and covers approximately 22,000 m$^{2}$ \citep{Albert_2020,Smith_2015}.   
HAWC is located at a latitude of $19^{\circ}$N, and current analyses can handle sources within $45^{\circ}$ of zenith.  Requiring some transit time within this range, and enough margin to form a map around the nova 
restricts HAWC's view of the sky to a declination range of about $+61^{\circ}$ to $-23^{\circ}$. This eliminates all but one of the 10 novae detected by \Flat between 2015 (when HAWC began operation) and 2019.  
V392 Per is located within HAWC's sky coverage and had a clear \Flat detection 
\citep{Li_etal18}.

V392 Per was discovered to be in eruption on 
2018 April 29 via the optical observations of amateur astronomer Yuji Nakamura \citep{cbat18,Endoh18}, and was later confirmed to be a Galactic nova by \citep{Wagneretal_2018}. V392 Per is located in the Galactic plane, but opposite the Galactic center (RA $= 70.8390^{\circ}$ and Dec $= 47.35719^{\circ}$ and in Galactic Coordinates $l = 157.9918^{\circ}$ and $b = 0.9022^{\circ}$).  This region has no strong TeV steady sources, which means that for HAWC, background estimation at this location  does not require subtraction of other sources. A geometric distance to V392~Per has been estimated by \citet{Chomiuk+21rad} to be
$3.5^{+0.7}_{-0.5}$\,kpc, using Gaia Early Data Release 3 \citep{Gaia16, Gaia21} and the prior suggested by \citet{Schaefer_2018}.
We use this distance in the
remainder of the paper.

\section{\Flat Observations of V392~Per}\label{sec:fermi}

GeV $\gamma$ rays were observed from V392~Per on 2018 April 30 at 6$\sigma$ significance with \Flat \citep{Li_etal18}, but no follow-up analysis of the nova's $\gamma$-ray behavior has yet been published. Here we analyze the \Flat light curve and spectral energy distribution (SED) of V392~Per. 

We downloaded the LAT data (Pass 8, Release 3, Version 2 with the instrument response functions of \texttt{P8R3\_SOURCE\_V2}) from the data server at the \textit{Fermi Science Support Center} (FSSC). 
The observations cover the period of 2018 Apr 30 to 2018 May 31 (note that there are no usable LAT data available for V392~Per between 2018 Apr 4--30 due to a solar panel issue).
For data reduction and analysis, we used \texttt{fermitools} (version 1.0.5) with \texttt{fermitools-data} (version 0.17)\footnote{\url{https://fermi.gsfc.nasa.gov/ssc/data/analysis/software} \\
and \url{/documentation/Pass8_usage.html}}. For data selection, we used a region of interest $14^\circ$ on each side, centered on the nova.
Events with the class \texttt{evclass=128} (i.e., SOURCE class) and the type \texttt{evtype=3} (i.e., reconstructed tracks FRONT and BACK) were selected. We excluded events with zenith angles larger than $90^\circ$ to avoid contamination from the Earth's limb. The selected events also had to be taken during good time intervals, which fulfils the \texttt{gtmktime} filter \texttt{(DATA\_QUAL$>$0)\&\&(LAT\_CONFIG==1)}. 

\begin{figure}[t!]
\begin{center}
\includegraphics[width=6.5 cm,
angle=90]{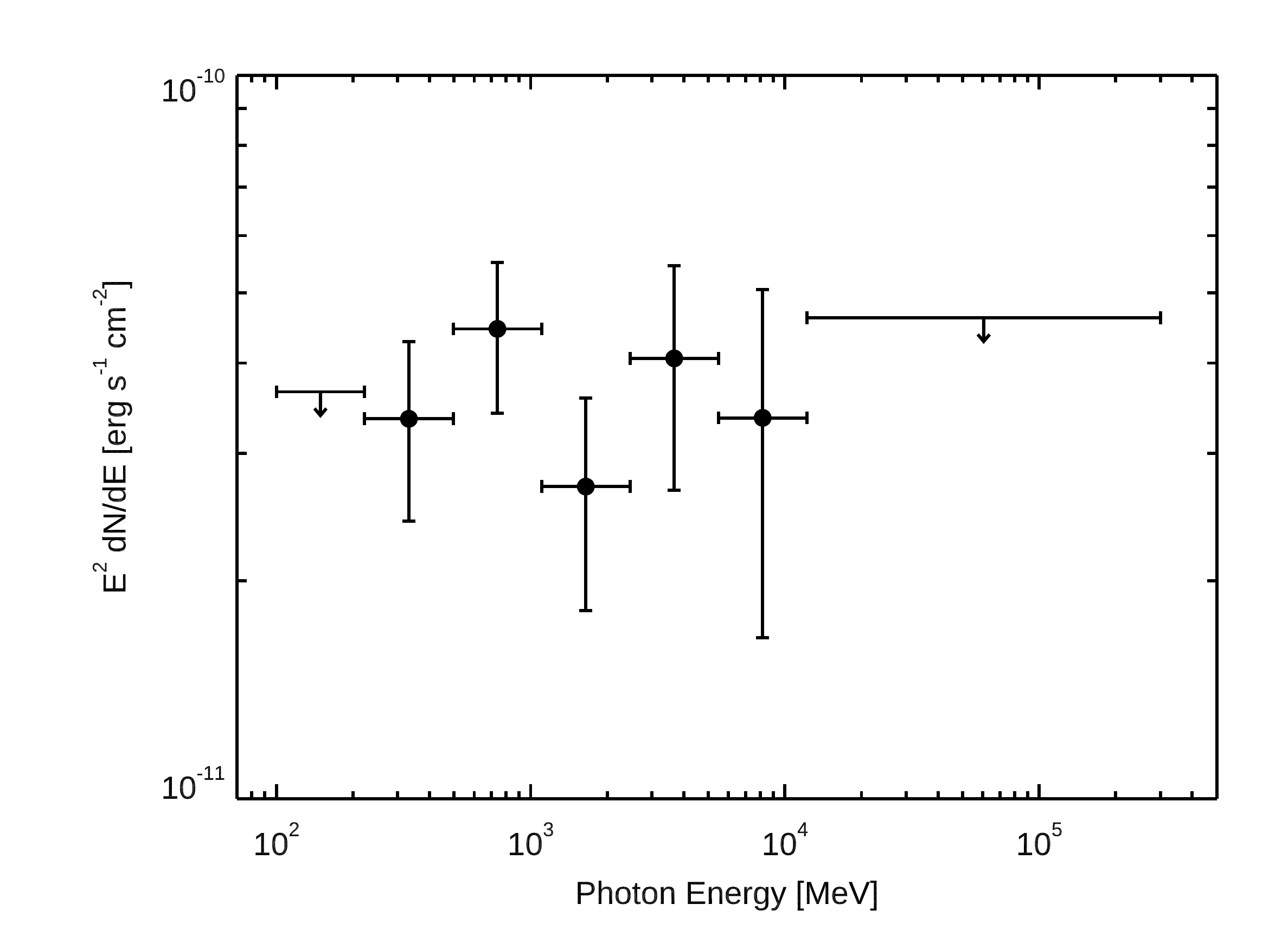}
\caption{The \Flat spectral energy distribution for V392 Per.  Upper limits in the lowest and highest energy bins signify 95\% confidence limits.}
\label{fig:fermi_sed}
\end{center}
\end{figure}

\begin{figure}[t!]
\begin{center}
\includegraphics[width=8.5 cm, angle=90]{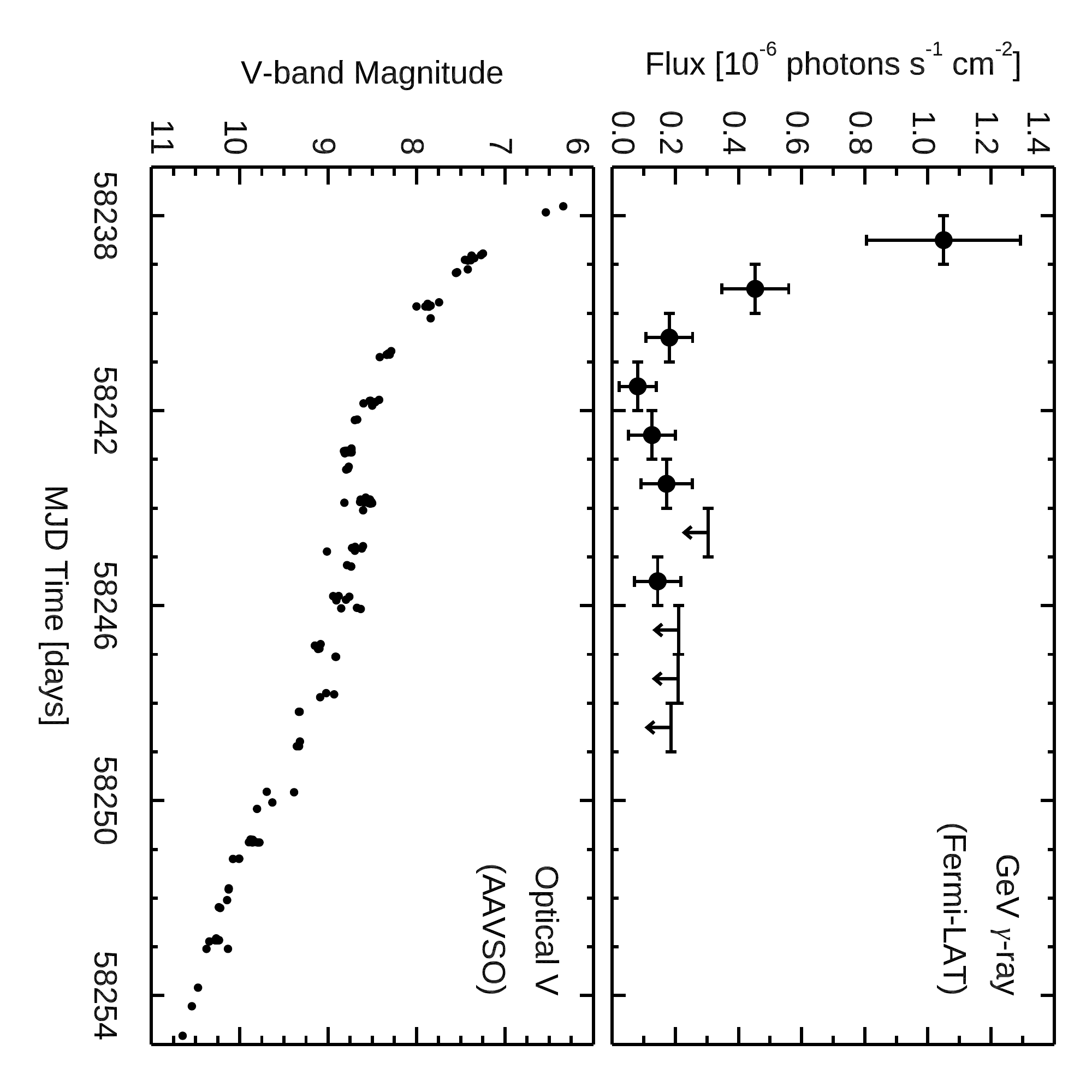}
\caption{\emph{Top:} the \Flat light curve for V392 Per. Photon flux is calculated over the energy range 0.1--300 GeV. The upper limits plotted are 95\% confidence limits. \emph{Bottom:} The optical $V$-band light curve of V392~Per over the same window of time as measured by AAVSO.}
\label{fig:fermi_lc}
\end{center}
\end{figure}

Next, we performed binned likelihood analysis on the selected LAT data. A $\gamma$-ray emission model for the whole region of interest was built using all of the 4FGL cataloged sources located within $20^\circ$ of the nova \citep{Abdollahi_etal20}. 
Since V392~Per is the dominant $\gamma$-ray source within 5$^\circ$ of the field, we fixed all the spectral parameters of the nearby sources to the 4FGL cataloged values for simplicity. In addition, the Galactic diffuse emission and the extragalactic isotropic diffuse emission were included by using the Pass 8 background models \texttt{gll\_iem\_v07.fits} and \texttt{iso\_P8R3\_SOURCE\_V2\_v1.txt}, respectively, of which the normalizations were allowed to vary during the fitting process. The spectral model of V392~Per was assumed to be a simple power law (PL)  model:
\begin{equation}
    \frac{dN}{dE} \propto E^{-\Gamma}.
\end{equation}
A preliminary light curve was first extracted with a spectral index 
$\Gamma=2$ (fixed) to investigate the $\gamma$ ray active interval. Using a $>2\sigma$ detection significance as a threshold (i.e., TS = $2\ \ln(L_{s+b}/L_b)\ >4$, where $L$ is the Poisson Likelihood Function) , we define the $\gamma$ ray active phase as eight days starting from 2018 Apr 30 (MJD 58238) to 2018 May 8 (MJD 58246). A stacked analysis in this period gives a detection significance of 11.6$\sigma$ (i.e., $TS = 133$). The average $\gamma$-ray flux integrated over 100~MeV--100~GeV over the \Flat detection period  is $(2.30 \pm 0.42) \times 10^{-10}$ erg s$^{-1}$ cm$^{-2}$ or $(2.19 \pm 0.41) \times 10^{-7}$ photons s$^{-1}$ cm$^{-2}$.
 A power law fit to the SED yields a best-fit photon index of $\Gamma = 2.0 \pm 0.1$ 
{ (coincidentally the same as the initially assumed $\Gamma = 2$) }
 and normalization, $F_\nu = (2.23 \pm 0.58) \times 10^{-9}$ photons s$^{-1}$ cm$^{-2}$ MeV $^{-1}$ at 100 MeV. 
The updated spectral model was then used to rebuild the \Flat light curve of V392~Per, which is plotted in Figure \ref{fig:fermi_lc}.
The GeV $\gamma$-ray spectral energy distribution (SED) of V392~Per is plotted in Figure \ref{fig:fermi_sed}.
Due to the limited data quality, we did not test other more complicated spectral models in the analysis (e.g., PL with exponential cutoff).

\section{HAWC Data Reduction and Analysis}

\subsection{Data Reduction}

HAWC is sensitive to $\gamma$ rays with energy above 300 GeV.  Based on the timing and locations of the PMTs struck by the shower, we reconstruct the location on the sky of the particle that initiated the shower.  For this analysis we use Right Ascension and Declination for the J2000 Epoch \citep{albert_et_al_2020_3hwc}. A key parameter for this analysis is \fhit, the fraction of PMTs that are struck during the shower event. This quantity can be used to parameterize the angular resolution and the $\gamma$-hadron selection criteria, and is sensitive to the energy of the initiating particle as described in \citep{albert_et_al_2020_3hwc} and \citep{Abeysekara_2017a,Abeysekara_2017b}.

In the remainder of this section, we show the statistical significance of HAWC observations, and report best fit flux and confidence limits (CL) assuming unbroken power laws. In \S 5 we set limits on the maximum (TeV) energy to which the \Flat SED could extend and be compatible with HAWC data; this method is applied to the nova for the first time, to our knowledge.  In \S 6 we provide HAWC limits in differing bins of true energy, without imposing the assumption of an unbroken power law as the SED shape; this method is new, to the best of our knowledge.  In \S \ref{sec:unc} we assess systematic uncertainties on the HAWC limits.  

\subsection{Results Assuming Simple Power Laws: Significance Maps, Best Fits, and Limits}

The time frame chosen for the main HAWC investigation of V392 Per covers 40 days, beginning 7 days before the optical discovery of the nova. 

For each day of the observation, we made a significance map of the region of interest for each of 9 bins of \fhit as described in \citep{albert_et_al_2020_3hwc}. Throughout this paper, we measure significance in units of standard deviations ($\sigma$).   
The same shocks that produce GeV $\gamma$ rays are also generally expected to be the source of any TeV radiation, so we analyzed HAWC data assuming the same $\Gamma = 2.0$ PL index as observed for the \Flat data.

\begin{figure}[htb]
    \centering
    \includegraphics[trim={.3cm 0 .3cm 0},clip,width= 9
        cm]{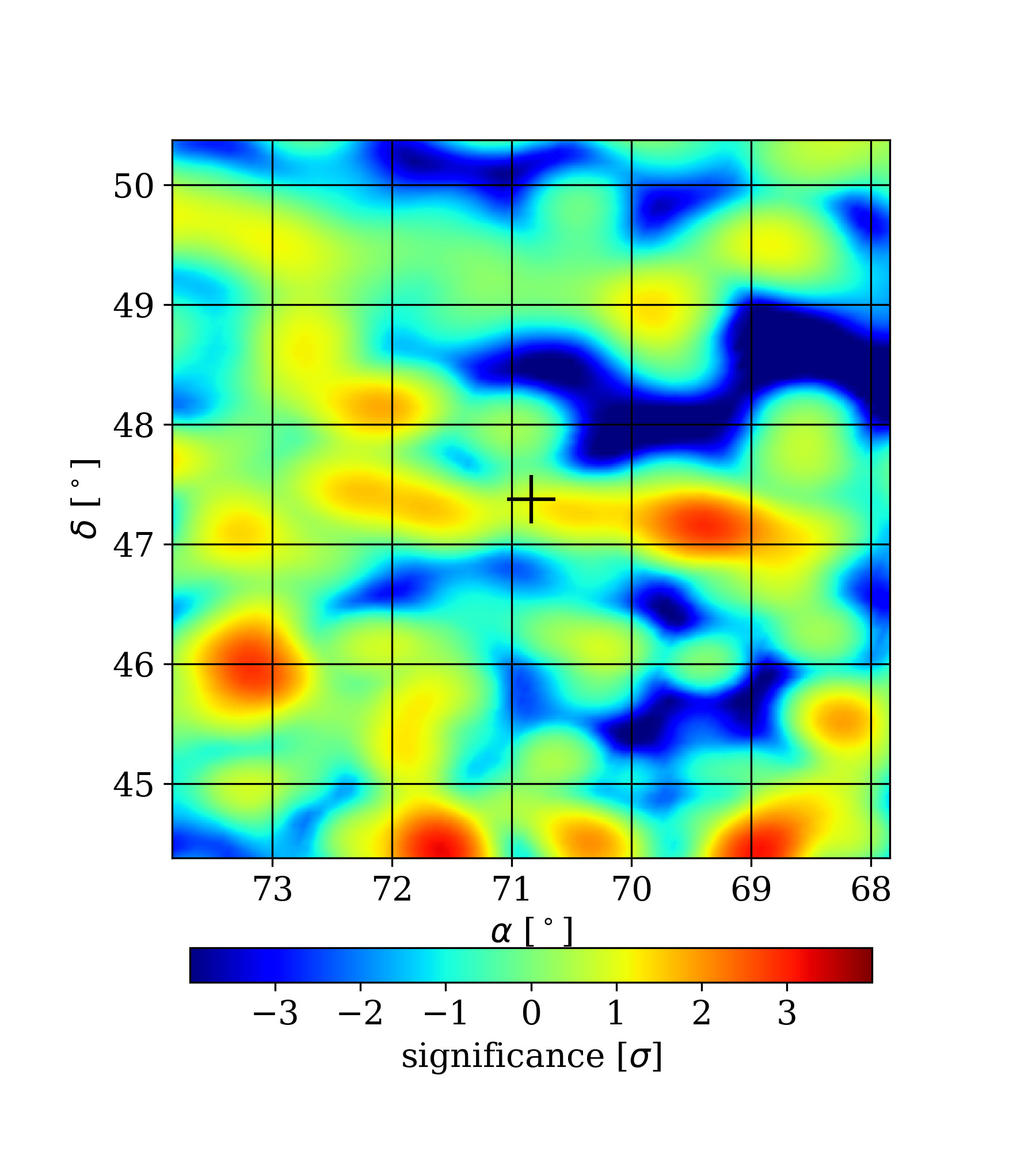}
    \caption{HAWC significance map of V392 Per during the week of \Flat detection (i.e., the ``On" period).  The position of the nova is indicated by a cross.  HAWC pointing at this declination is accurate to better than 0.1 degree.}
    \label{fig:7_Day_Sig}
\end{figure}
 
We define three periods within this time range.  The ``On" period covered 7 days starting 2018 April 30  (MJD 58238 to 58245), the same as the \Flat 8-day active period except for the last day, when we had power issues at the HAWC site.  The ``Before" period is 7 days starting 2018 April 23 (MJD 58231 to 58238), before the ``On" period. This includes one day of optical activity during which \Flat was not observing due to a solar panel problem.   The ``After" period is 7 days after the end of the ``On" period, starting on 2018 May 8 (MJD 58246 to 58253).
In addition, we defined a 7 day ``On$- 1$ yr" period on the same days as the ``On" period, but a year before V392 Per's eruption, in order to represent a period when no signal is expected. 

For each period, we performed forward-folded fits of a $\Gamma=2.0$ PL spectral model to the 9 HAWC pixel-level data maps for each \fhit bin, centered at the V392 Per location as in \citep{albert_et_al_2020_3hwc}.
Throughout the rest of the paper we report best fit values for the SED point at E = 1 TeV ($S=E^2\ dN/dE$), its uncertainty (dS) or corresponding 95\% upper Confidence Limits (CL) (\ $S_{95}$); all are in units of erg s$^{-1}$ cm$^{-2}$. Results are for the ``On" period whenever no specific period is given. We use the method described in \citep{Albert_2018} for setting 95\% CL.  The SED points and SED 95\% CL values in this paper were calculated using the  HAL 
\footnote{https://tinyurl.com/2p93m3tz (threeml hal\_example)}
(HAWC Accelerated Likelihood) plugin
\citep{brisbois21hal,younk2015highlevel} to the 3ML multi-mission analysis framework \citep{vianello:3ml.2015}.

We also calculate the statistical significance of the normalization of the $\Gamma=2.0$ PL SED compared to zero TeV emission. A significance map is this calculation as a function of sky position. Figure \ref{fig:7_Day_Sig} shows the significance map 
during the ``On" period contemporaneous with the \Flat GeV detection.  There is a mild excess of 1.6$\sigma$ significance near the nova location.  

\begin{figure}[htb]
    \centering
    \includegraphics[width= 8.5cm
    ]{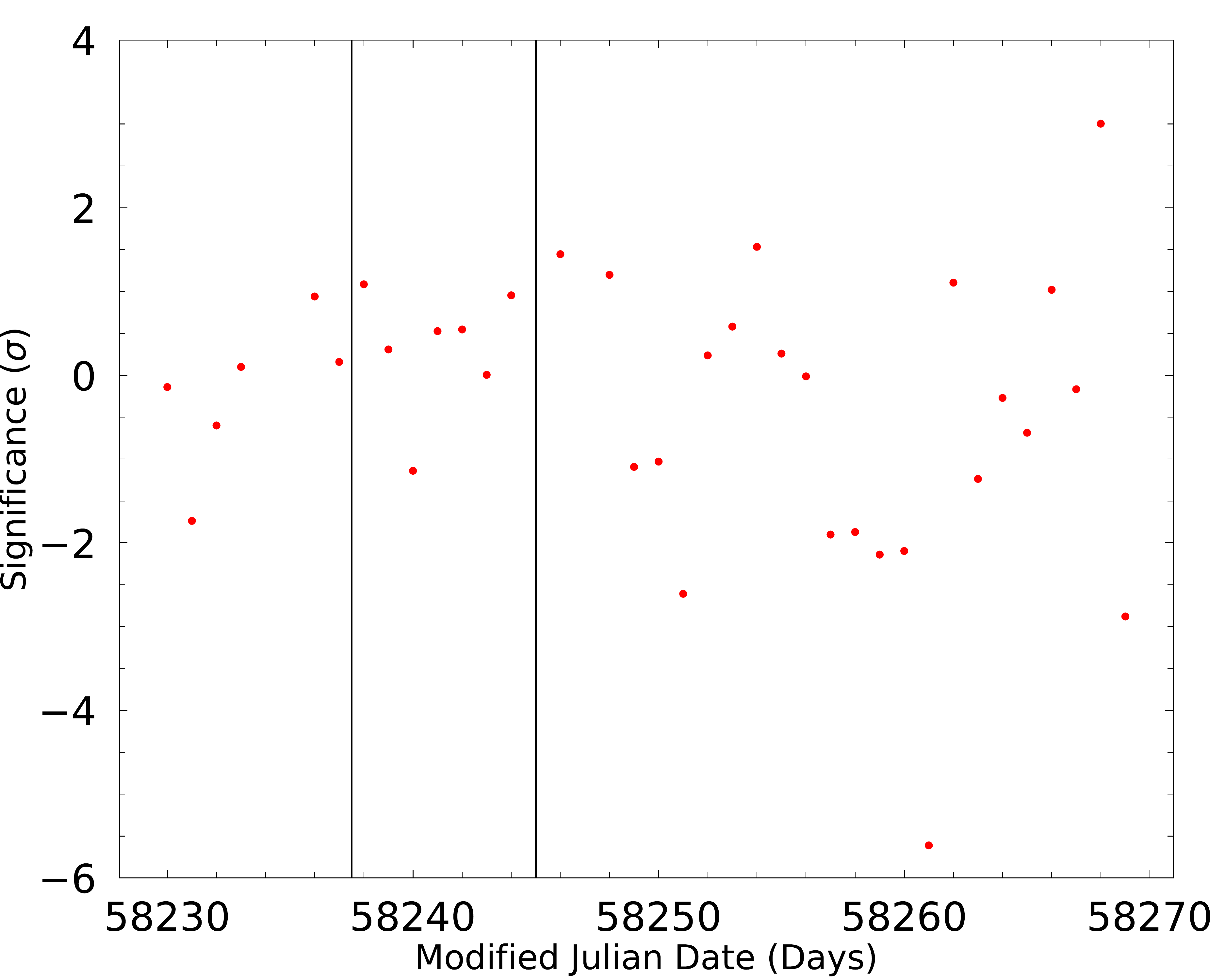}
    \caption{Daily significance of V392 Per from 2018 Apr 22 to 2018 May 31 in HAWC data, assuming a $\Gamma=2.0$ PL spectrum.  The vertical lines bound the ``On" period of observation, coincident with the \Flat GeV detection. }
    \label{fig:Daily Significance 2018_1}
\end{figure}

Figure \ref{fig:Daily Significance 2018_1} shows the significance at the nova position for each day during the study period, with the ``On" period indicated between the black lines.  Some transits are missing when electrical storms or power outages interfered with HAWC data taking.

Table \ref{tabs:plfits} shows limits from a HAWC SED fit and the resulting significance for a $\Gamma = 2.0$ PL spectral model for all the time periods, as well as the best \Flat SED fit.  While there is a weak $1.6\ \sigma$ suggestion of TeV emission during the ``On" period, and an even weaker hint during the ``After" period, the best fit HAWC flux and 95\% upper limit on the flux is far less than would be expected for a continuation into the HAWC TeV regime of the $\Gamma = 2.0$ PL seen by \Flat in the GeV regime. 
\begin{table}[htb]
\begin{tabular}{lrrrr}
\toprule
 & $S $ & $dS$ & $S_{95}$ & $Z_0$\\
\toprule
``On" & 1.2 & 1.1  & 3.9  & 1.6\\
``Before" & $-$1.9  & 1.1  & 1.4  & $-1.7$\\
``After" & 0.5  & 1.2  & 2.7  & 0.5\\
``On$-1$ year" & $-0.1$  & 0.7  & 1.4  & $-0.2$\\ 
\Flat & 35  & 10 & & 11.6\\
\bottomrule
\end{tabular}
\caption{Best fit SED point at 1 TeV ($S$) in units of $10^{-12}$ erg s$^{-1}$ cm$^{-2}$, its uncertainty ($dS$), and the 95\% upper limit on the SED point ($S_{95}$), assuming a PL spectral model with $\Gamma =2.0$  for HAWC data over three time periods and for the \Flat data.  Also shown is $Z_0$, the statistical significance of the observation in standard deviations.   Negative best fit fluxes occur half the time when no real source exists.  We also show for comparison the results of the \Flat SED fit in the GeV range described in \S 3.
}
\label{tabs:plfits}
\end{table}

\begin{figure}[htb]
    \centering
\includegraphics[width=8.5cm,  ]{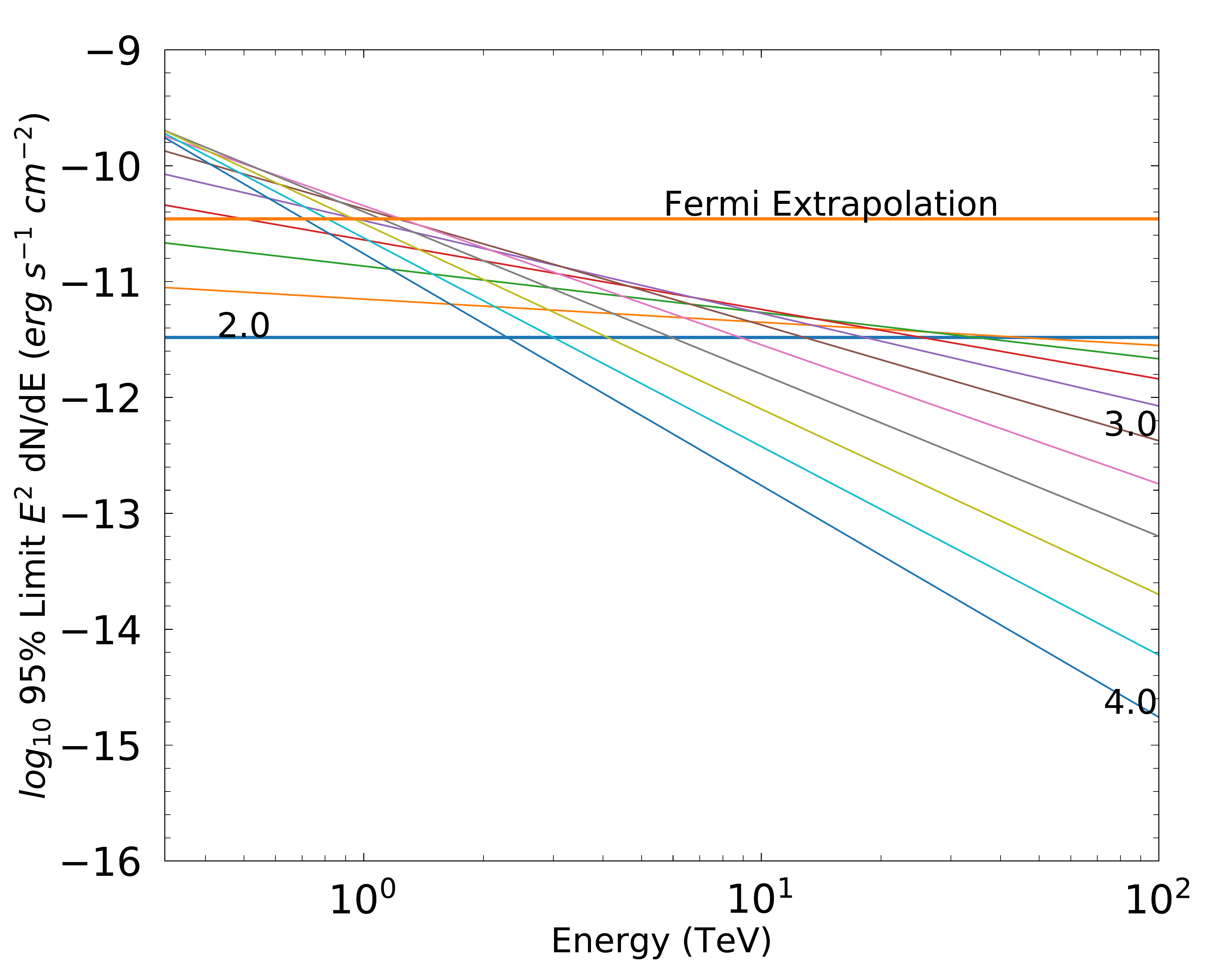}
    \caption{HAWC upper limits on the SED corresponding to $S_{95}$ during the ``On" period, varying the PL index between $\Gamma = 2-4$ in steps of 0.2 (limits for 2.0, 3.0, and 4.0 are noted specifically). The upper horizontal line, in orange, denotes the \Flat best fit SED extrapolated to high energy.
    }
    \label{fig:indices}
\end{figure}

Next we considered the effect of changing the PL index. Figure \ref{fig:indices} shows the SEDs corresponding to the $S_{95}$ limit for various PL indices.  Also shown is the best fit to the \Flat flux assuming an unbroken PL extending to very high energies.  Softer PLs (larger indices) produce less restrictive limits at low energy.  The upper envelope of the lines in Figure \ref{fig:indices} can be thought of as a SED limit as a function of energy, independent of the actual value of the PL index---at least within the family of PL spectrum shapes \citep{pooja}.   All the limits are statistical only; we discuss systematic uncertainties in \S \ref{sec:unc}.

\section{Hypothesis Tests for Maximum TeV Detected Energy} \label{hypoth}
We now quantify the level at which a \Flat SED extension to TeV energies is disfavored by the HAWC data.  
{In the null hypothesis $H_0$, we constrain the PL normalization to that found by \Flat.  For the alternative hypothesis $H_1$ we take the normalization from a best fit to HAWC data.  In a series of hypothesis tests, we use a $\Gamma = 2.0$ PL spectrum model  with a step-function cutoff at some maximum energy.  While this spectrum ends too abruptly to describe an actual nova spectrum, it allows us to consider the evidence against having observed TeV photons from V392 Per above a given energy. }

{We  calculate the significance (in standard deviations) of the disagreement of observed HAWC data flux with the \Flat SED extended to the cutoff energy by
\begin{equation}
Z_F = (S_F - S)/dS_F
\end{equation}
where $S_F$ is the \Flat flux from the last row of Table \ref{tabs:plfits}, $S$ is the best fit HAWC flux to the cut-off spectrum, and $dS_F$ is the uncertainty of a measurement of a simulated source with the strength of the \Flat flux, again for the cutoff hypothesis spectrum.}
  The results are shown in Table \ref{tabs:hyptest}.  
We also show $Z_0$, the number of standard deviations by which the best fit flux is favored over no TeV emission at all, and show the best fit flux values for each assumed cutoff. 

\begin{table}[htb]
\centering
{\begin{tabular}{ccrrr}\toprule
Cutoff E & $S$ &$Z_0$& $dS_F$ & $Z_F$ \\
(TeV) &  & ($\sigma$) & & ($\sigma$)\\
\toprule
5  & 4.1  &  0.3 & 10.4 & 1.7\\
10 & 6.5 & 1.0 & 5.2 & 2.8\\
15 & 6.0 & 1.4 & 4.2 & 3.7\\
\bottomrule
\end{tabular}
}
\caption{Hypothesis test of a $\Gamma = 2.0$ PL with various hard cutoffs at high energy.  The flux translated into an SED point ($S$) is the best fit of a $\Gamma = 2.0$ PL with the specified cutoff energy to HAWC data,in units of $10^{-12}$ erg s$^{-1}$ cm$^{-2}$; 
{in the same units, $dS_F$ is the uncertainty of a fit to an injected source with the \Flat flux $S_F$ (35 in these units, taken from Table \ref{tabs:plfits}). } $Z_0$ is the significance of the HAWC flux (compared to zero flux).
 $Z_F$ represents the significance by which the HAWC best fit SED differs from an extension of the \Flat GeV SED to the cutoff energy.
}
\label{tabs:hyptest}
\end{table}

 The best-fit SED is always more than a factor of 5 below the \Flat extension SED.   The HAWC data  
 {reject (by $Z_F$ near 3 standard deviations, or more)} extension of the 2.0 PL to 10 TeV or higher.  Emission below 5 TeV at the extension flux level is not as strongly excluded.  This is because HAWC is more sensitive at higher energies, as we will discuss further in the next section.  All the truncated spectra fit to HAWC data have  a significance ($Z_0$) less than two standard deviations compared to zero flux.

\section{TeV Flux Limits as a Function of Photon Energy}

\subsection{HAWC Flux Limits in Energy Bins}
\begin{figure}[!b]
    \centering
   \includegraphics[width= 8.5cm, ]{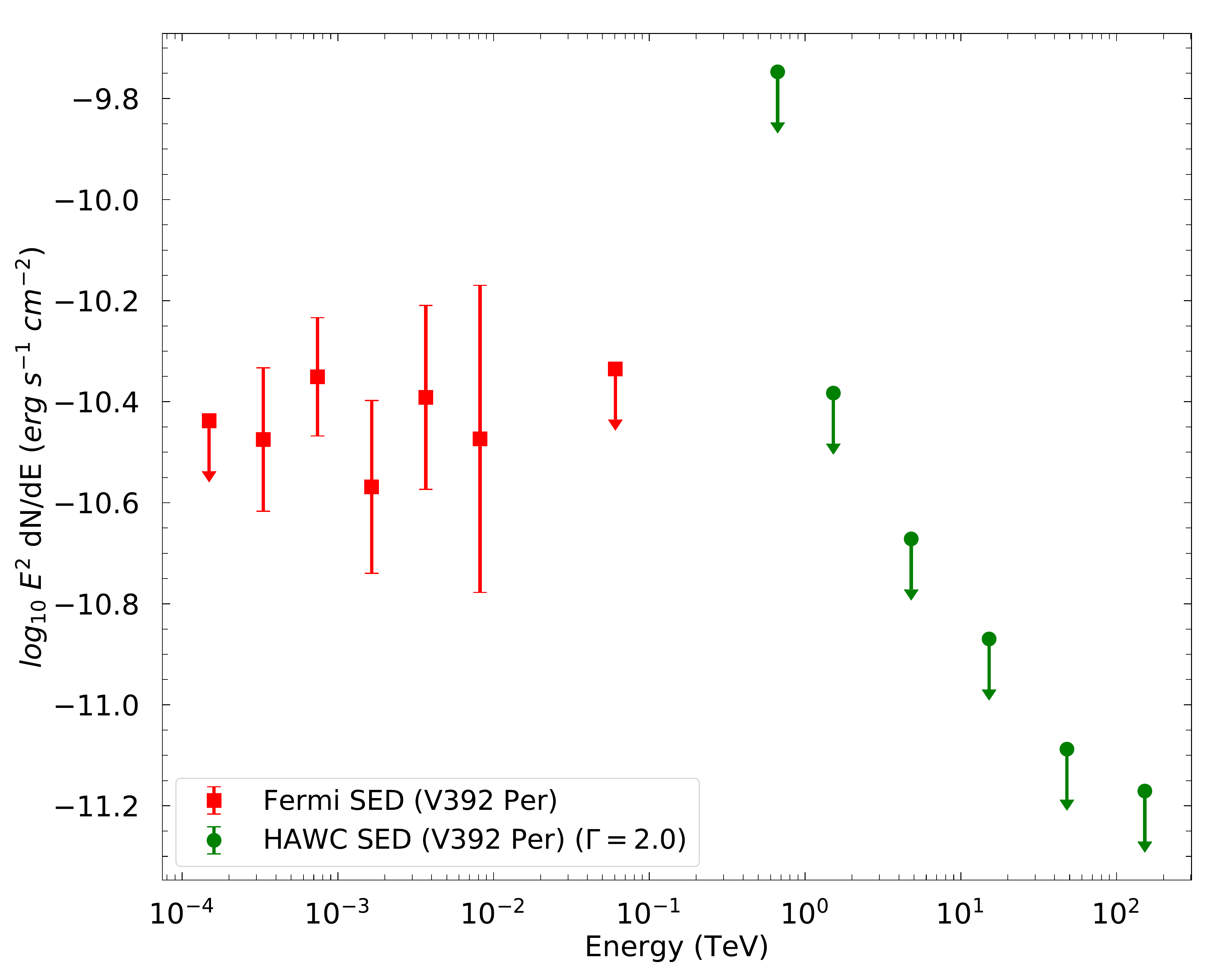}
    \caption{The $\gamma$-ray SED of V392 Per, plotting \Flat detections and HAWC 95\% upper limits.  The $\Gamma=2.0$ refers to a local power law assumed within a HAWC energy bin in the limit calculation.}
    \label{fig:difflimits}
\end{figure}

We now present limits in bins of energy, assuming a $\Gamma = 2.0$ PL index within each energy bin.  
In Figure \ref{fig:difflimits} we show  the \Flat SED for V392 Per and the $S_{95}$ HAWC upper limits. 
This analysis uses maps binned in \fhit, and its energy resolution effects are reasonably matched by half-decade energy bins (e.g. 1--3.16, 3.16--10 TeV, etc.). HAWC energy estimators could provide better resolution at higher energy, but the additional event selection criteria would reduce sensitivity to a transient source such as a nova. 

The method used to find limits in true energy bins, using data consisting of maps binned in \fhit, is as follows.  First, we perform a forward-folded fit of the energy spectrum assumed, a point source model, and the detector response including the point spread function to the set of data maps for only the normalization $\hat{k}$ of a PL of form, $E^{-2}$\ where E is in TeV.  Then for each true energy bin $j$ we perform a second fit for the normalization $k_j$ of the $\Gamma = 2.0$ PL, but with the contribution of energy bin $j$ removed from the original unrestricted power law, in a way that retains the best fit contributions of all other energy bins as determined by $\hat{k}$ from the original fit.  Specifically, we fit to the data the form
\begin{equation}
    dN/dE = \hat{k}\ \{\ E^{-2} -  bin(E,j)\ E^{-2}\ \} + k_j\ bin(E,j)\  E^{-2}
\end{equation}
where $bin(E,j) = 1 $ when $E$ falls between the lower and upper edges of energy bin $j$.  Finally, the normalization of the limit is determined by increasing the value of $k_j$ until the fit log likelihood increases by an amount (2.71/2) appropriate for a 95\% confidence limit.  This method allows us to report a limit separately in each individual energy bin, without assuming an overall PL SED, as the normalization of each energy bin is determined separately.  Because now each energy bin contains less data than the whole of the data, these limits are, however, less than those in \S 4, where a single unbroken PL is assumed for the underlying SED.

\begin{figure}[htb]
    \centering
    \includegraphics[width= 8.5cm, ]{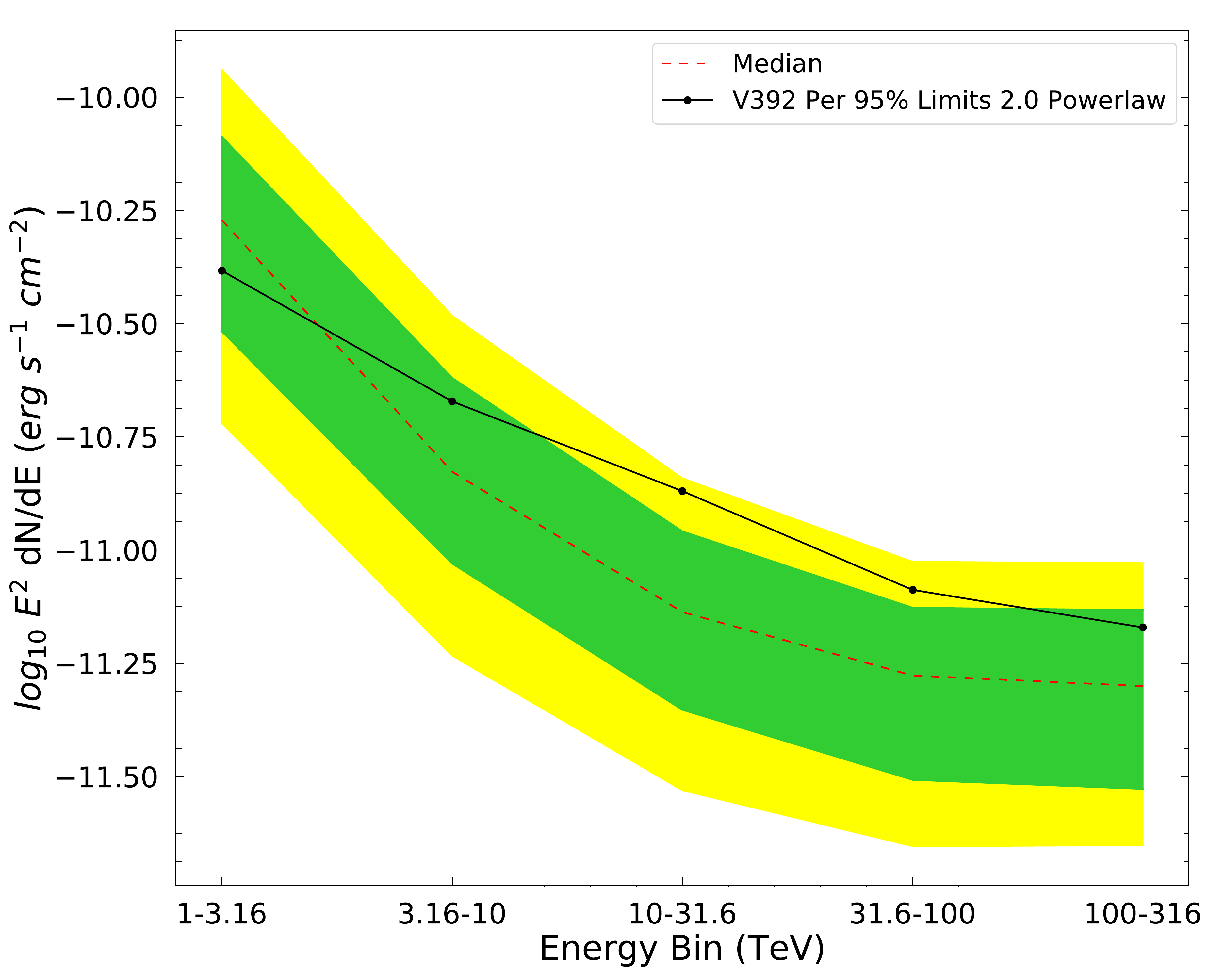}
    \caption{Expected and Observed Limits vs. Energy for V392 Per.  The observed limit is in black and the median expected limit is a dashed line in red.  The central green band covers 68\% of expected limits (1 $\sigma$) while the outer yellow bands cover 95\% (2 $\sigma$).}
    \label{fig:BBvsEnergy}
\end{figure}

The limit from the lowest energy HAWC bin is compatible with continuation of the \Flat SED, but higher energy bins are incompatible at the 95\% CL.  The limit from fitting a single $\Gamma = 2.0$ PL across the full HAWC energy range is considerably more restrictive, placing a 95\% upper limit at $E^2 \frac{dN}{dE} = 4.0 \times 10^{-12}$ erg s$^{-1}$ cm$^{-2}$ (Figure \ref{fig:indices}).

We simulated by Monte Carlo the expectations for the energy-dependent limits for each energy bin under the hypothesis of no physical flux (only Poisson fluctuations of the background).  The distribution of expected limits is shown in Figure \ref{fig:BBvsEnergy}.  The inner (green) and outer (yellow) bands covers 68\% and 95\% of the simulated limits respectively and the central dashed (red) line shows the median of the expected limits in each energy bin.  The observed limits (from Figure \ref{fig:difflimits}) are shown here in a black solid line to allow comparison with the expected distribution of limits assuming no flux. The observed limits for bins above 3 TeV are typically 1-2 standard deviations above expectation,  consistent with either a modest statistical fluctuation or weak TeV emission.

\subsection{Comparison with other TeV Nova Limits}

There have been two previous TeV observations of novae detected by \Flat.  Both observations were made by imaging air Cherenkov telescopes (IACTs).  IACTs have better point-source sensitivity  than HAWC, but IACTs can only observe sources which fall into their limited field of view (a few degrees). This typically requires specific pointing, a source visible at night, and good weather. As a result, it is harder for IACTs to observe contemporaneously with a \Flat observation.  In contrast, HAWC observes 2/3 of the sky daily.

 \begin{figure}[!htb]
    \centering
    \includegraphics[width= 8.5cm, ]{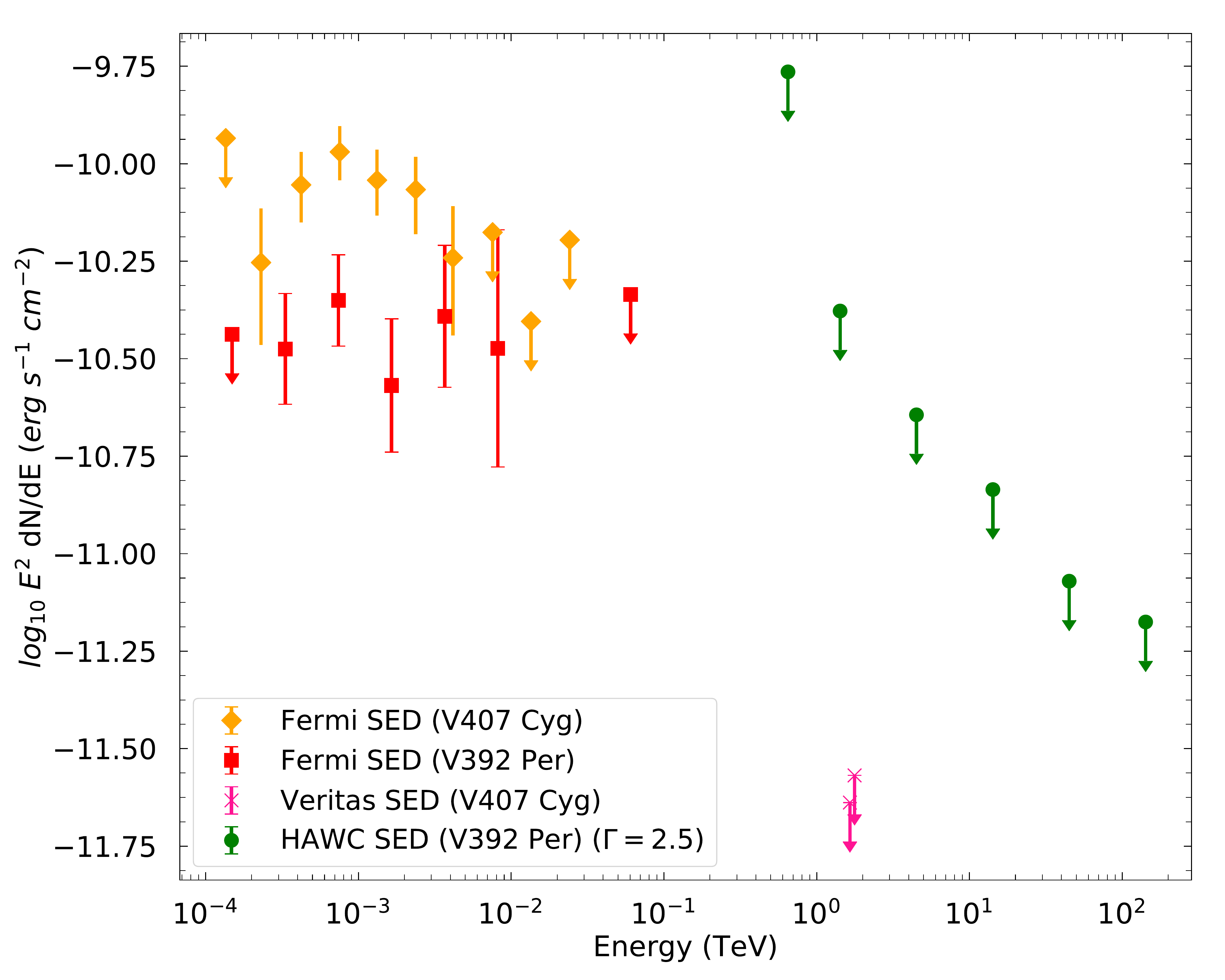}
    \caption{The $\gamma$-ray SED of nova V407 Cyg, with \Flat detections in yellow and VERITAS limits in magenta. Also plotted is the SED for V392 Per, with HAWC limits  using the same $\Gamma= 2.5$ PL index as VERITAS.}
    \label{fig:veritas}
\end{figure}

The first search for TeV nova emission was by the VERITAS collaboration on V407 Cyg \citep{Aliu_2012}.
VERITAS began observations 9 days after the beginning of the \Flat detection, and extended over a week of continued \Flat detection. VERITAS was unable to detect significant flux above 0.1 TeV, and set 95\% limits as shown in Figure \ref{fig:veritas}.  
Figure \ref{fig:veritas} also shows the \Flat SED as reported in \citep{Abdo_etal10}.  
Because of the curvature of the \Flat SED, VERITAS analyzed their data with a $\Gamma = 2.5$ PL.    The VERITAS limit is quoted at energies of 1.6--1.8 TeV, where the limit and assumed PL are least correlated (depending slightly on which of two analysis methods were used). To roughly compare with HAWC sensitivity, the HAWC differential limits on V392 Per are shown, but re-analyzed with the same $\Gamma = 2.5$ PL.
These limits are calculated as described above, but instead of $\Gamma = 2.0$, using $\Gamma = 2.5$.

\begin{figure}[!htb]
    \centering
    \includegraphics[width= 8.5cm, ]{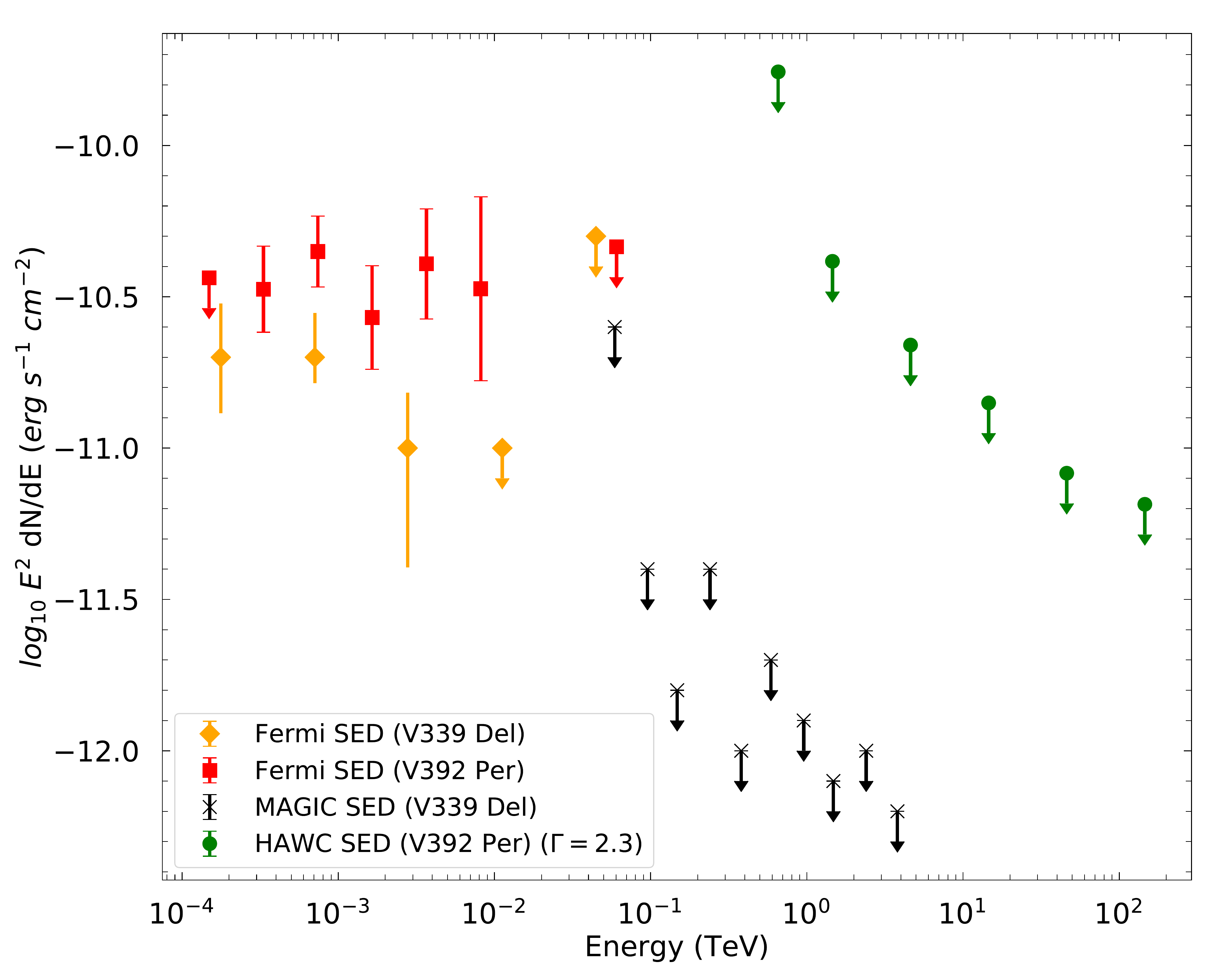}
    \caption{The $\gamma$-ray SED of nova V339 Del, with \Flat detections in yellow and and MAGIC limits in black. Superimposed is the SED of V392 Per (red and green), with HAWC limits for V392 Per using the same $\Gamma= 2.3$ PL index as MAGIC.}
    \label{fig:magic}
\end{figure}

MAGIC searched for TeV emission from V339 Del \citep{Ahnen_etal15}, which was slightly fainter than V392 Per in the GeV band \citep{Ackermann+14}. They found no TeV detected flux and produced limits shown in Figure \ref{fig:magic}.  The MAGIC analysis used a $\Gamma = 2.3$ PL index, motivated by a fit to the observed \Flat SED. 
Again for rough comparison, we show HAWC's V392 Per limits analyzed with this $\Gamma = 2.3$ PL.  At the overlapping energies, the MAGIC results were about 30 times more constraining than our HAWC limits.  
It is also worth mentioning that MAGIC was able to observe one night at the beginning of the nova's GeV $\gamma$-ray  detection, albeit under poor conditions; that observation produced a flux limit about a factor of 10 worse than their best nights of observation 9--12 days later---by which time the GeV $\gamma$-ray signal had faded, though not as much as V392 Per had faded by its second week. 

Thus, previous IACT nova observations produced stronger constraints on TeV emission than HAWC, and started from lower energy than HAWC limits. However, they only apply to the period 9 days after the beginning of the optical nova; HAWC's observations began 2 days after the optical nova, and temporally overlap the entire period of GeV detection with \Flat.  The HAWC ``After" period (days 9--15 of the optical nova) matches the time delay of the VERITAS and MAGIC observations.  Table \ref{tabs:plfits} suggests the ``After" period places slightly more restrictive limits than for the ``On" period.

\section{Systematic Uncertainties in HAWC Analysis}
\label{sec:unc}

\begin{table}[!htb]
\centering
{\begin{tabular}{lcc}\toprule
Effect & \% $-$change & \% +change\\
\toprule
late light&$-2$ & 8\\
charge&$-5$ & 2\\
threshold&$-2$ & 2\\
response&$-5$&-\\
\bottomrule
combined&$-7$\%&9\%\\
\end{tabular}
}
\caption{Systematic uncertainties in $S_{95}$, the 95\% CL SED  at 1 TeV from a $\Gamma = 2.0$ PL spectral model, and their combination in quadrature.
}
\label{tabs:syst}
\end{table}
Here we list the main systematic uncertainties affecting the HAWC results.  These uncertainties reflect discrepancies between data and events from the HAWC detector simulation as discussed in \citep{Abeysekara_2019} and \citep{albert_et_al_2020_3hwc}.  The size of the effects in this analysis will differ from those described in these references, because the analyses undertaken are different. We quantify their effects by the changes in $S_{95}$ at 1 TeV from the $\Gamma = 2.0$ PL spectral model in the ``On" period. The size of each effect is given in Table \ref{tabs:syst}; when relevant, we show the possible impact in both a possible increase (+change) or decrease ($-$change) in $S_{95}$.   We estimate the size of an effect by running our analysis using a plausible alternative detector response and comparing the result with our best-estimate detector response. 

\textit{Late light.--}
This effect comes from the fact that the laser light used in the calibration system has a narrower time distribution than the arrival of light from air shower events.  This is one of the largest sources of uncertainty.

\textit{Charge Uncertainty.--}
This encapsulates differences in relative photon efficiency among PMTs, and the uncertainty of PMT response to a given amount of Cherenkov radiation. 

\textit{Threshold uncertainty.--}
The PMT threshold is the lowest charge our PMT  electronics can register; despite studies, it is imperfectly known. It is the smallest among the main uncertainties.

\textit{Detector response parameterization.--}
The baseline detector response used is the same as in \citep{albert_et_al_2020_3hwc}.  This detector response was simulated for
declination values spaced by  one degree, so the best-match declination is quite close
to that of V392 Per.  Overall, this is judged to be the best available response file.  However, this response was calculated using weighting
(within \fhit bins, and for parameterization of the point spread function) for a 
$\Gamma = 2.63$ PL, while we typically fit a $\Gamma =2.0$ PL.  We considered an alternative detector
response  calculated with a $\Gamma =2.0$ PL weighting, but which had been evaluated every 5
degrees of declination (coarser than ideal as some of our software selects the
best declination match to a source, rather than interpolating).  Our estimate of the effect of the uncertainty in detector response is the difference between $S_{95}$ for these two response files, neither of which is ideal.

The systematic uncertainties are summarized in Table \ref{tabs:syst}.  Because the
effects are independent of each other, we separately combine in quadrature the positive
and negative effects.  The net result is that our limits carry approximately 8\%
systematic uncertainty in either direction.

\section{Modeling of V392 Per}
Before modeling the gamma-ray emission from V392 Per we need to understand first the environment surrounding the nova. 
In \S \ref{sec:bol}, optical photometry is used to estimate the bolometric flux of V392 Per as a function of time after the outburst. In \S\ref{sec:spec}, we use optical  measurements of the H$\alpha$  $(n=3 \rightarrow n=2)$  line profile to estimate the velocity of the slow and fast flows in the ejecta (and the resulting shock).   
In \S \ref{sec:ex}, we use optical spectra taken 6 days after $t_0$, to measure absorption from the interstellar medium along the line of sight, and measure the resulting extinction from the associated dust column.
The bolometric flux values are corrected for dust extinction, and when combined with the \emph{Gaia} distance measurement, the bolometric luminosity is calculated.

We then describe 
the $\gamma$-ray emission from V392 Per. Collisions 
among nova ejecta shells, or between the ejecta and an external environment, form shocks which accelerate ions to relativistic energies. These relativistic particles collide with surrounding gas to produce pions, which then decay into $\gamma$-ray photons observable by \Flat and potentially, HAWC. In \S \ref{sec:GeVprops}, we place the GeV properties of V392 Per in context of other $\gamma$-ray detected novae.
We next consider our ability to observe TeV photons, as they are limited by absorption due to $e^{\pm}$ pair creation, which depends on the density of optical photons the TeV photons must pass through (\S \ref{sec:tau}). This radiation density depends on the nova luminosity, the radius of the shock, and the spectral shape of the optical emission.
In \S \ref{sec:emax}, the nova's bolometric luminosity and shock velocity are used to estimate the magnetic field in the shock region, which in turn determines the maximum energy of the accelerated particles and hence of their $\gamma$-ray emission.    
\subsection{Optical Input Parameters} \label{sec:optical}
The modeling of V392 Per's $\gamma$-ray emission requires input parameters derived from optical data.  Here we show how we derived these values.

\begin{figure*}[!t]
\begin{center}
  \includegraphics[width=8.5cm]{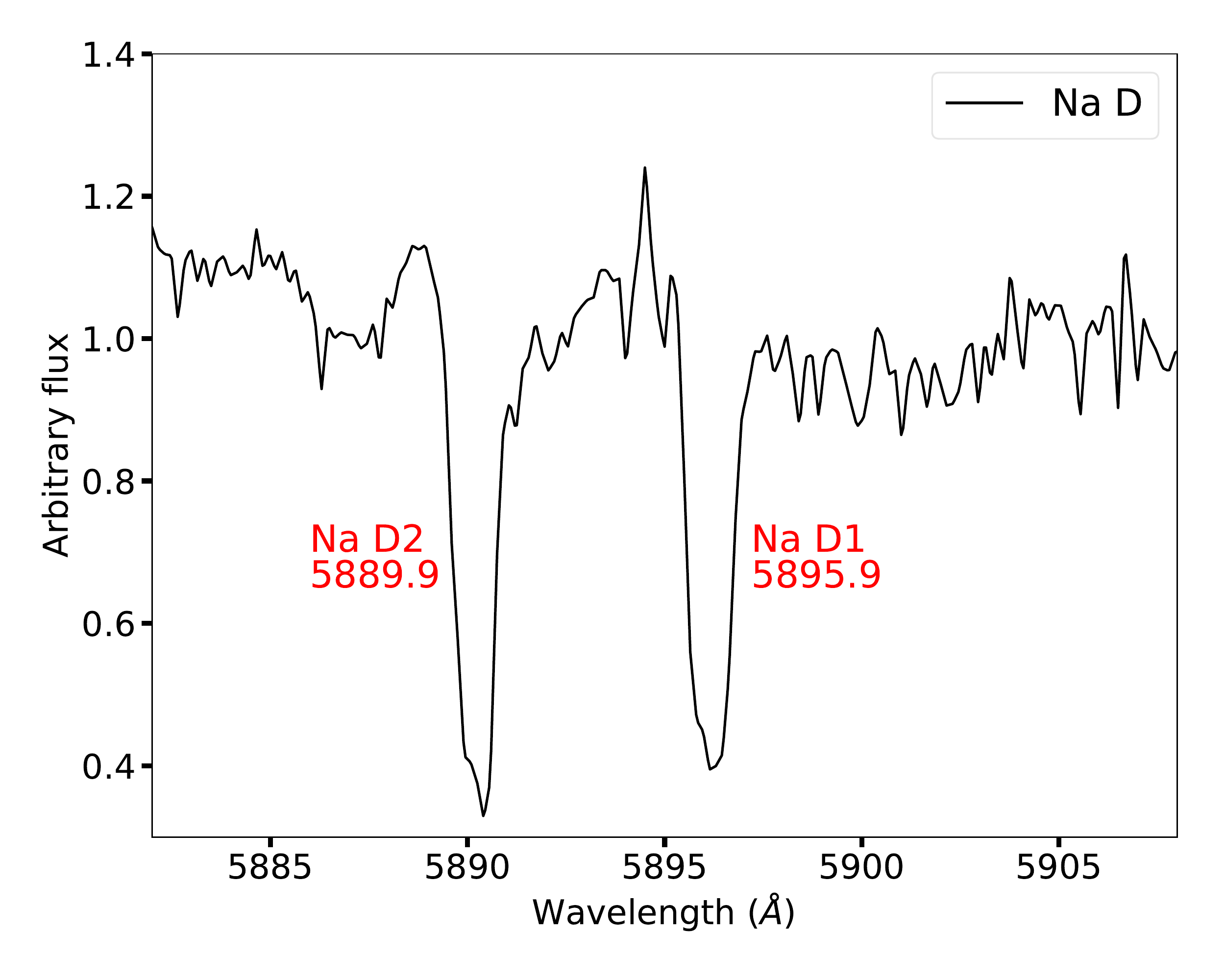}
  \includegraphics[width=8.5cm]{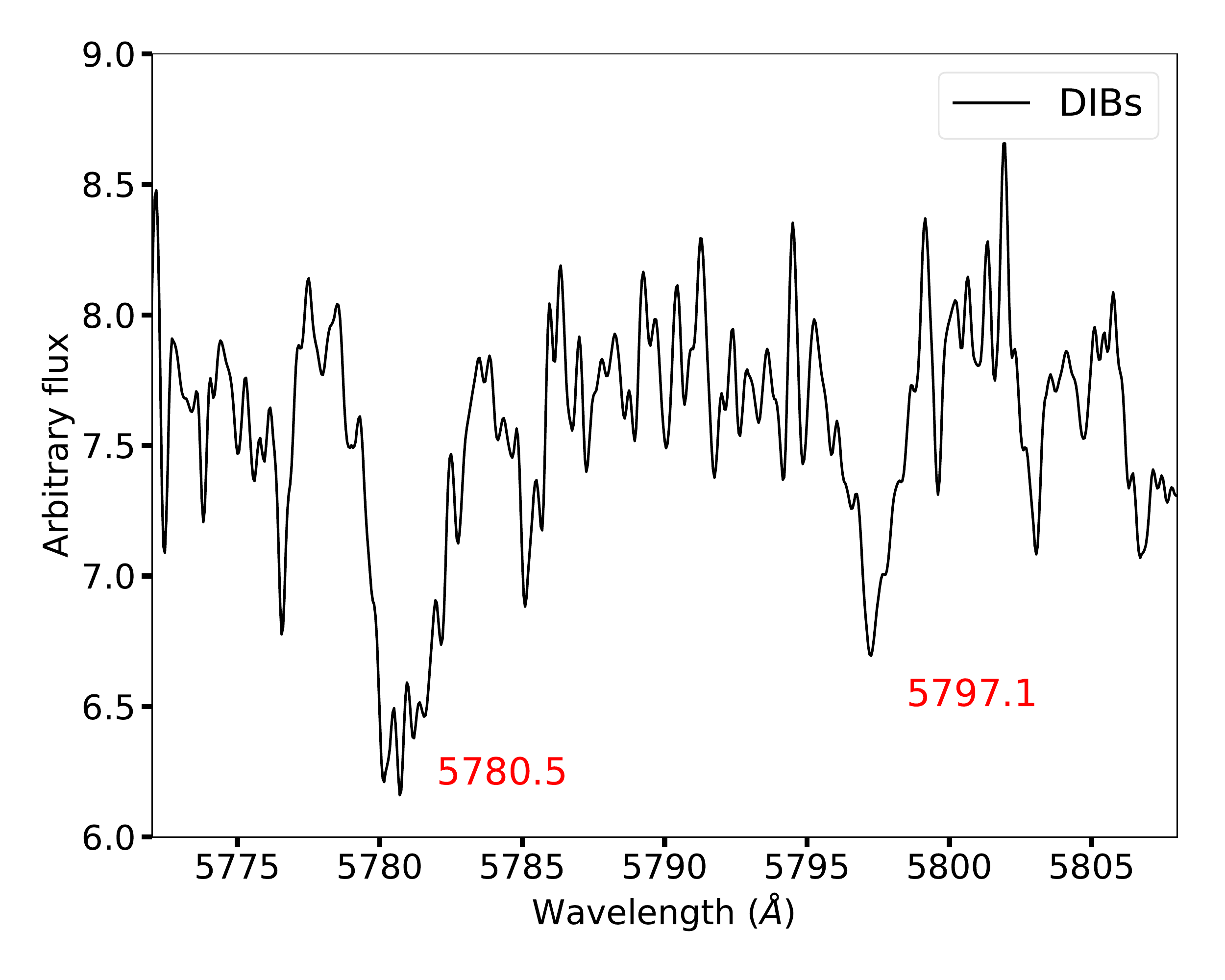}
\caption{The spectral lines used to estimate the reddening taken 6 days after $t_0$. \textit{Left}: the Na~I D interstellar absorption lines at 5895.9\,\AA  (D1) and
5889.9\,\AA (D2). \textit{Right}: the diffuse interstellar bands
used to estimate the reddening.}
\label{Fig:DIBs}
\end{center}
\end{figure*}

\subsubsection{Extinction from Interstellar Dust}  \label{sec:ex}

To estimate the extinction due to interstellar dust along the line of sight to V392 Per, we rely on several interstellar absorption lines: the Na~I D doublet and some diffuse interstellar bands (Figure~\ref{Fig:DIBs}). In this and \S \ref{sec:spec}, we make use of publicly available spectra from the Astronomical Ring for Access to
Spectroscopy (ARAS\footnote{\url{http://www.astrosurf.com/aras/Aras_DataBase/Novae.htm}}; \citealt{Teyssier_2019}). The low- and
medium-resolution spectra cover the first month of the optical outburst, starting
from 
{the time of optical maximum} (day zero). 
{To measure the interstellar lines, we used a high-resolution spectrum from day 6.} 

Based on the equivalent widths of the Na~I D lines and the empirical relations of \citet{Poznanski_etal_2012}, we derive a reddening value, $E(B-V) = 0.78 \pm 0.04$  mag, to V392~Per. Based on the equivalent width of the two absorption lines at
5780.5 and 5797.1\,\AA\  and the empirical relations from
\citet{Friedman_etal_2011}, we derive $E(B-V) = 1.04 \pm 0.05$ mag. This leads to an
average reddening value of $E(B-V) = 0.90 \pm 0.18$ mag for V392 Per. For a standard interstellar extinction law ($A_V = 3.1\, E(B-V)$; e.g., \citealt{Mathis90}), we find a $V$-band extinction value,  $A_V = 2.8 \pm 0.5$ mag.
This value is consistent with the one derived by 
\citet{Chochol+21}, and we
use it in the remainder of the paper.

\subsubsection{Bolometric Luminosity}\label{sec:bol}
Multi-band optical photometry was performed by several observers from the American
Association of Variable Star Observers (AAVSO; \citealt{Kafka_2020}) from day zero (April 29, 2018; 
{the time of discovery of eruption and also the time of optical maximum})
and throughout the optical outburst of V392~Per 
{(see \citealt{Chochol+21} for a more detailed description of the light curve)}. We make use of photometry in the $BVRI$ bands to estimate the nova's total (bolometric) luminosity in the few weeks following the nova eruption.
Near optical peak, the optical pseudo-photosphere of the nova reaches its maximum radius and the SED is characterized by an effective temperature of 6000 - 10000 K,  peaking in the $BVRI$ bands (e.g.,
\citealt{Gallagher_Starrfield_1976,Hachisu_Kato_2004,Bode&Evans08}).

\begin{figure}
\begin{center}
  \includegraphics[width=8.5cm]{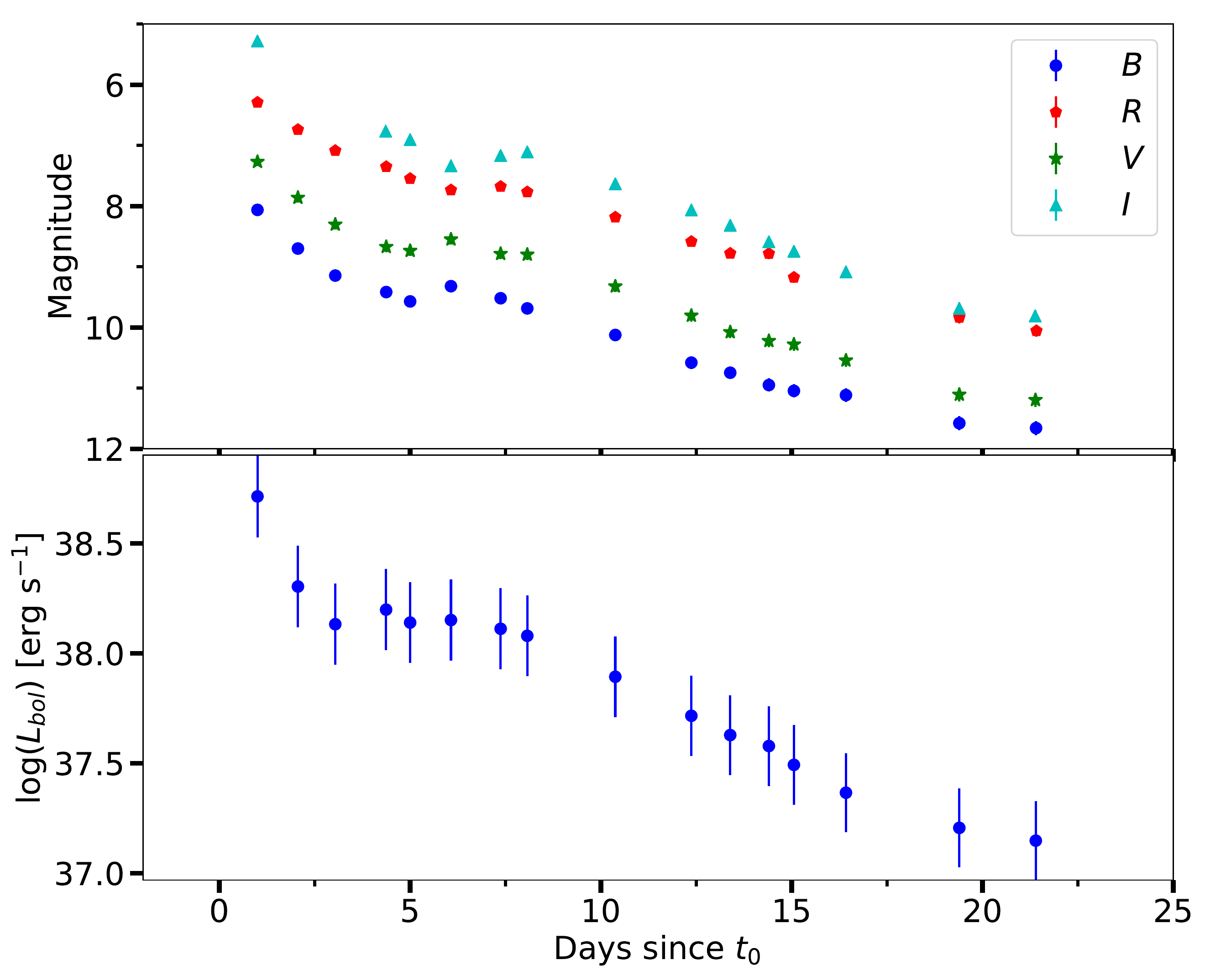}
\caption{\textit{Top}: optical light curve of V392~Per, measured in the  $B$, $V$, $R$, and $I$ bands by the AAVSO. \textit{Bottom}: V392 Per's bolometric luminosity as a function of time, estimated from integrating the optical SED.}
\label{Fig:bol}
\end{center}
\end{figure}

In order to estimate the bolometric luminosity of the nova as a function of time, we used the \verb|bolometric| task which is part of the SNooPy python package \citep{Burns+11}. This task directly integrates the flux measured by the $BVRI$ photometry (we used \verb|method=`direct'|), which adds a Rayleigh-Jeans extrapolation in the red (\verb|extrap_red=`RJ'|), and corrects this SED for extinction from intervening dust (we use $A_V$ = 2.8 mag; see Section~\ref{sec:ex}). We plot the $BVRI$ photometry, along with the derived bolometric luminosity, in Figure \ref{Fig:bol}.

\subsubsection{Expansion Velocities from Spectral Line Profiles}\label{sec:spec}

\begin{figure}
\begin{center}
  \includegraphics[width=8.5cm]{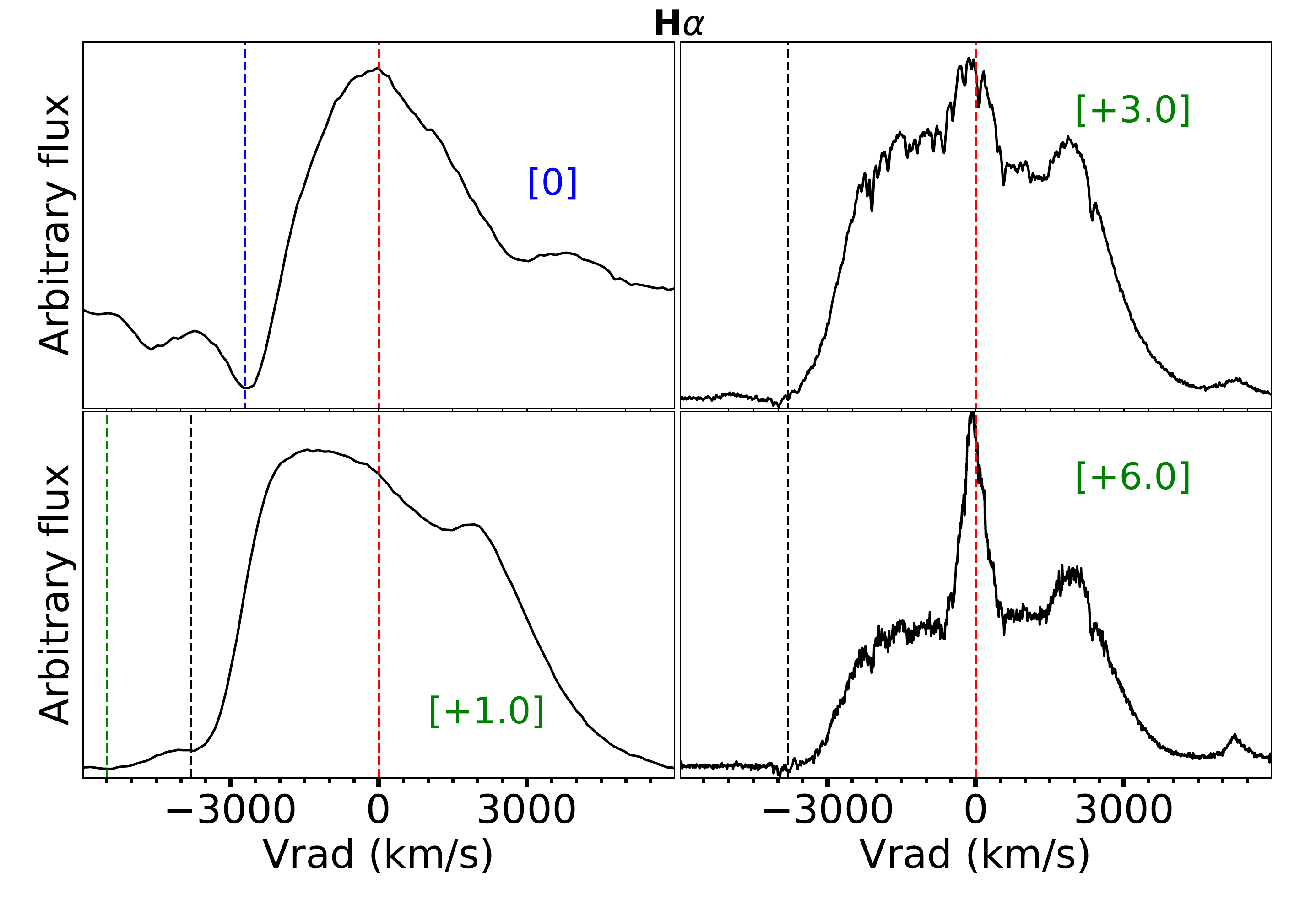}
    \includegraphics[width=8.5cm]{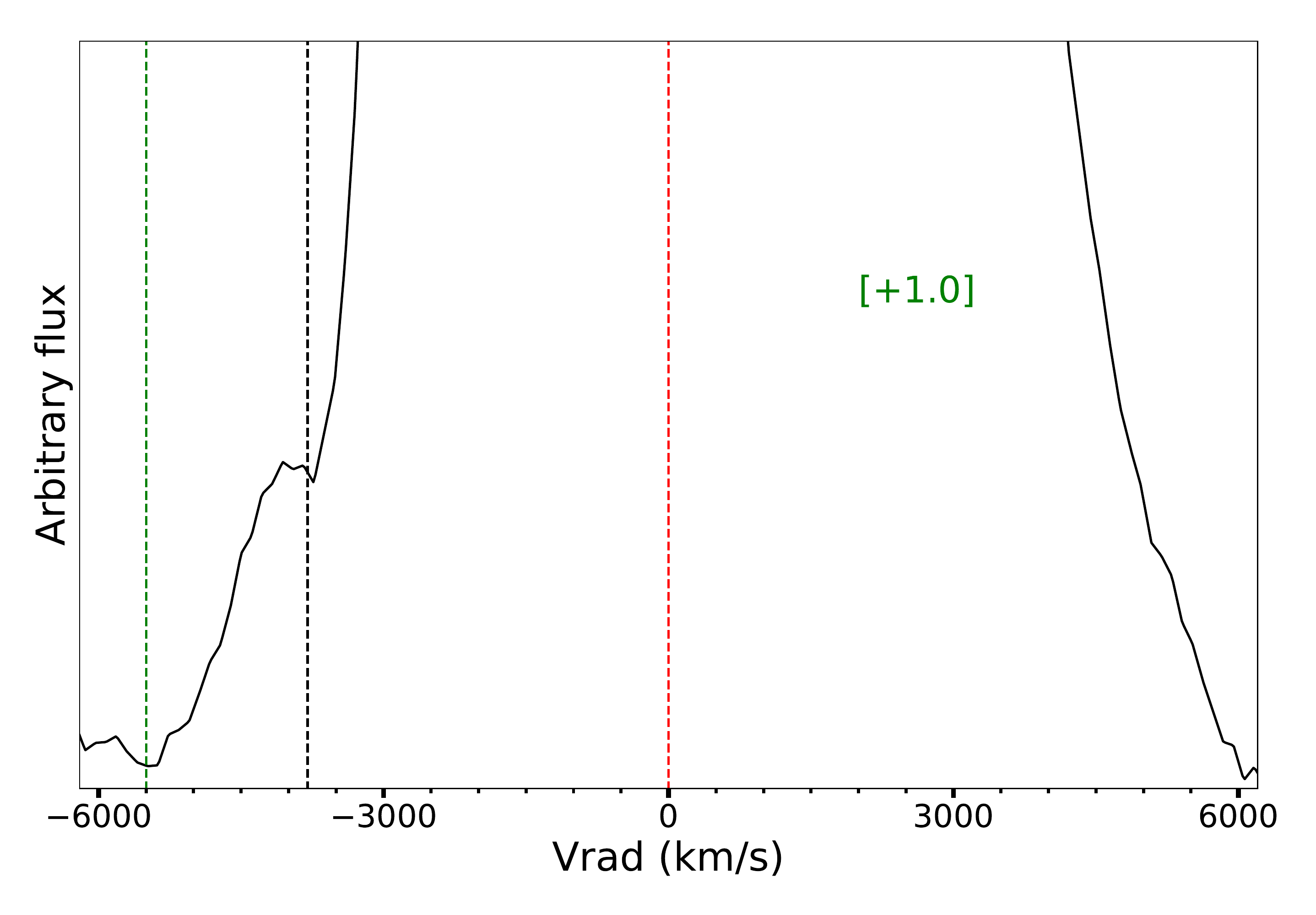}
\caption{\textit{Top:} the evolution of V392 Per's H$\alpha$ line profile near optical peak. Numbers in brackets represent days relative to the peak. The radial heliocentric velocities
are derived relative to the line center, which is marked by a red dashed line. The
blue, black, and green dashed lines represents $v_1$ = 2700\,km\,s$^{-1}$, $v_{2}$ =
3800\,km\,s$^{-1}$, and $v_3$ = 5500\,km\,s$^{-1}$ relative to the line center,
respectively. \textit{Bottom:} a zoom in on the absorption components for the day +1 spectrum.}
\label{Fig:line_profiles}
\end{center}
\end{figure}

Figure~\ref{Fig:line_profiles} shows the spectral
evolution of H$\alpha$ during the first few days of the eruption of V392~Per. 
{As noted by \cite{Wagner+18} and \cite{Chochol+21},} the spectral 
lines initially show a P Cygni profile with an absorption trough at a blueshifted
velocity of around $-$2700\,km\,s$^{-1}$ (blue line in Figure~\ref{Fig:line_profiles}). On day +1, a broader emission component
emerges, extending to blueshifted velocities of around $-$5500\,km\,s$^{-1}$ (green line in Figure~\ref{Fig:line_profiles}; see also the zoom-in on this profile in the rightmost panel of Figure~\ref{Fig:line_profiles}).
This indicates the presence of two
physically distinct outflows: a slow and a fast one, as described in \citet{Aydi_etal_2020b}. At this time, there is another absorption component, superimposed on the broad emission, with  a velocity of around 3800\,km\,s$^{-1}$ (black line in Figure~\ref{Fig:line_profiles}). This component, which appears around optical peak and has an intermediate velocity between the slow and fast component is the so-called ``principal component'' as historically classified by \citet{McLaughlin43,Mclaughlin47}.
\citet{Friedjung87} and \citet{Aydi_etal_2020b} suggest that this intermediate velocity component is the
outcome of the collision between the initial slow flow and the following faster
flow, and therefore the velocity of the intermediate component depicts the speed $v_{\rm cs}$ of
the cold central shell sandwiched between the forward and reverse shocks \citep{Metzger+14}. 

\subsection{GeV $\gamma$-ray behavior of V392 Per}\label{sec:GeVprops}

\Flat detections of V392~Per were only made for eight days following optical maximum, in one of the shortest duration and most sharply peaked $\gamma$-ray light curve yet observed from a nova (see Figure 8 of \citealt{Chomiuk+21araa}).  We note that the turn-on of the $\gamma$-rays was not fully captured in V392 Per, as \Flat was suffering technical problems during its rise to optical maximum, so this duration is a lower limit. However, the true duration is unlikely to be substantially longer than observed, given that \Flat signals tend to first become detectable around optical maximum \citep[e.g.,][]{Ackermann+14}, V392 Per's observed optical maximum was on 2018 April 29.8 \citep{Chochol+21}, and \Flat observations resumed on April 30. 

\begin{figure}
\begin{center}
  \includegraphics[width=8.5cm]{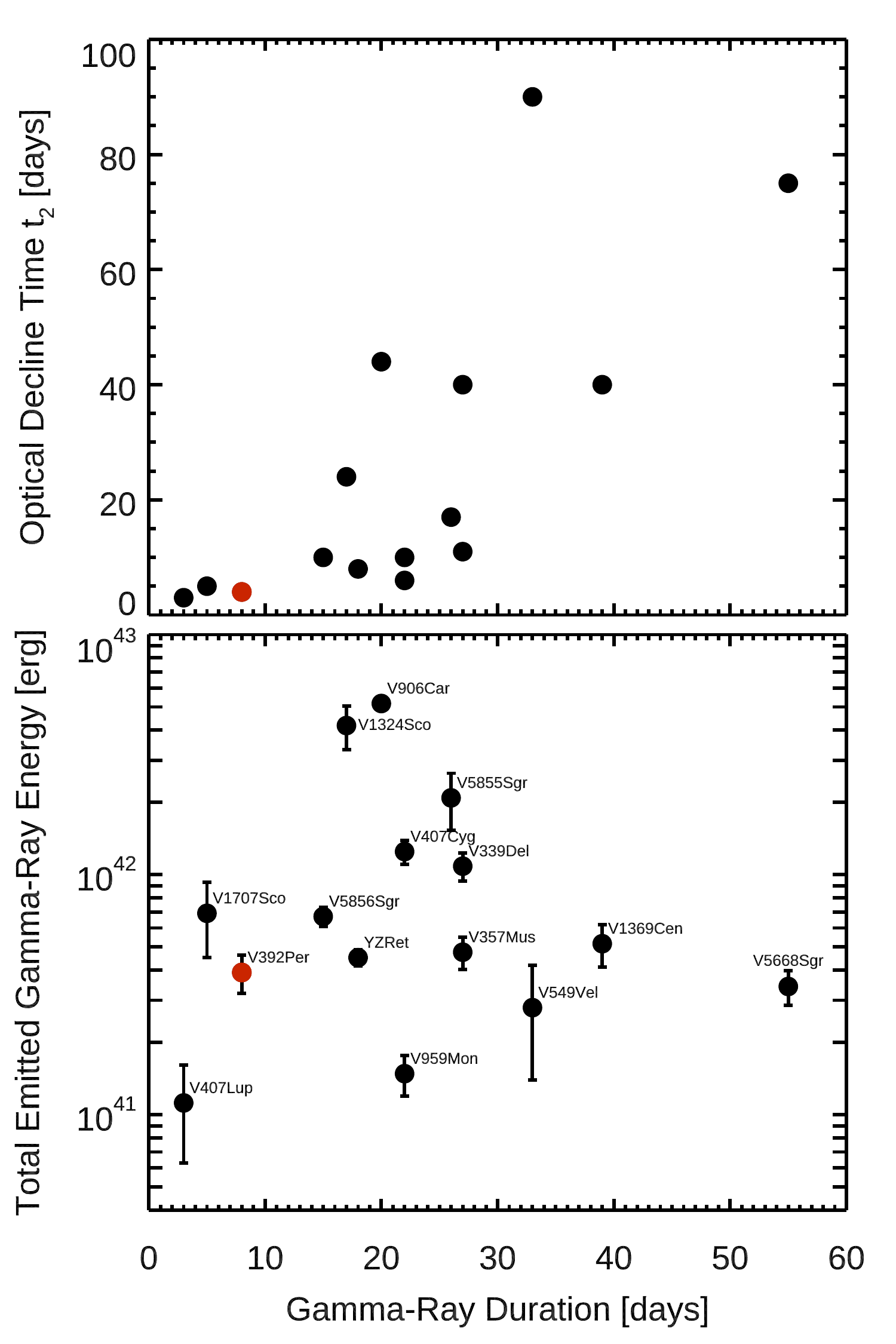}
\caption{\emph{Top:} For each of 15 $\gamma$-ray detected novae, the duration over which \Flat achieved $> 2 \sigma$ detections is plotted against the time for the optical light curve to decline by 2 magnitudes from maximum ($t_2$). \emph{Bottom:} $\gamma$-ray duration plotted against the total energy emitted during this time integrated over the LAT bandpass. In both panels, V392 Per is plotted as a red point.}
\label{Fig:fermidur}
\end{center}
\end{figure}

The short duration of the \emph{Fermi} signal in V392 Per is perhaps not surprising, as $\gamma$-ray light curves have been observed to correlate and covary with optical light curves in novae \citep{Li+17, Aydi+20}, and  V392~Per's optical light curve evolves very quickly (Figure \ref{fig:fermi_lc}). In the top panel of Figure \ref{Fig:fermidur}, we compare the duration of \Flat $\gamma$-rays against the time for the optical light curve to decline by two magnitudes from maximum ($t_2$) for the 15 $\gamma$-ray detected novae tabulated in Table S1 of \cite{Chomiuk+21araa} (see \citealt{Gordon+21} for associated $t_2$ values). 
{We see that  novae that are slower to decline from optical maximum generally remain $\gamma$-ray bright for longer. A Spearman rank correlation test gives $p = 0.0002$ (for a one-tailed test), indicating a significant correlation between the gamma-ray duration and the optical decline time.} With $t_2 = 5.9$ days, V392 Per has one of the fastest evolving optical light curves and a similarly rapid $\gamma$-ray light curve to match.

During its \Flat detection, the GeV $\gamma$-ray luminosity of V392~Per was on average $5 \times 10^{35}$ erg s$^{-1}$. Such a luminosity is typical amongst $\gamma$-ray detected novae, which show variations in \Flat luminosity of $>$2 orders of magnitude (see Figure S1 of \citealt{Chomiuk+21araa} along with \citealt{Franckowiak+18}). 
The average $\gamma$-ray luminosity but short duration of V392~Per motivated us to plot $\gamma$-ray duration against total energy emitted in the \Flat band in the bottom panel of Figure \ref{Fig:fermidur}, comparing V392 Per (in red) with data on fourteen other \emph{Fermi}-detected novae \citep{Chomiuk+21araa}. Based on \Flat light curves of five novae, \citet{Cheung+16} found a tentative anti-correlation between these properties, with the counter-intuitive implication that novae which remain $\gamma$-ray bright for longer emit less total energy in the \Flat band. Figure \ref{Fig:fermidur} revisits this claimed anti-correlation with three times the number of \emph{Fermi}-detected novae, and we find that it no longer holds; there are many novae with relatively short $\gamma$-ray duration and relatively low total $\gamma$-ray energy, with V392~Per among them.

\subsection{$\gamma$-ray attenuation in V392 Per}\label{sec:tau}
Before addressing the implications of the TeV non-detections by HAWC, we must ask whether such emission could even in principle be detected, due to absorption processes that occur close to the emission site at the shock.  Of particular importance at TeV energies is attenuation due to pair creation, $\gamma-\gamma\ \rightarrow e^{-} + e^{+}$, on the background radiation provided by the optical light of the nova.  In contrast, at the GeV energies that \Flat  is sensitive to, pair creation would require X-ray target photons. Attenuation is therefore less important in the GeV range than in the TeV range, because the X-ray luminosity (and photon number density) of novae is low compared to optical/UV during the early phases of nova eruptions when $\gamma$-ray emission is observed.  

Other forms of gamma-ray opacity, such as photo-nuclear absorption (Bethe-Heitler process), are comparatively less important than the $\gamma\gamma$ opacity.  In particular, the Bethe-Heitler opacity increases slowly with photon energy, being only a factor $\sim 3$ times larger at 100 TeV than at 1 GeV \citep{Chodorowski+92}; hence, if the Bethe-Heitler optical depth through the nova ejecta is low enough to permit the escape of gamma-rays detectable by {\it Fermi}-LAT, then it is unlikely to impede the escape of photons in the HAWC energy range across the same epoch, particularly considering that the optical depth of the expanding ejecta is expected to decrease rapidly with time.

\begin{figure}
    \centering
     \includegraphics[width=8.5cm]{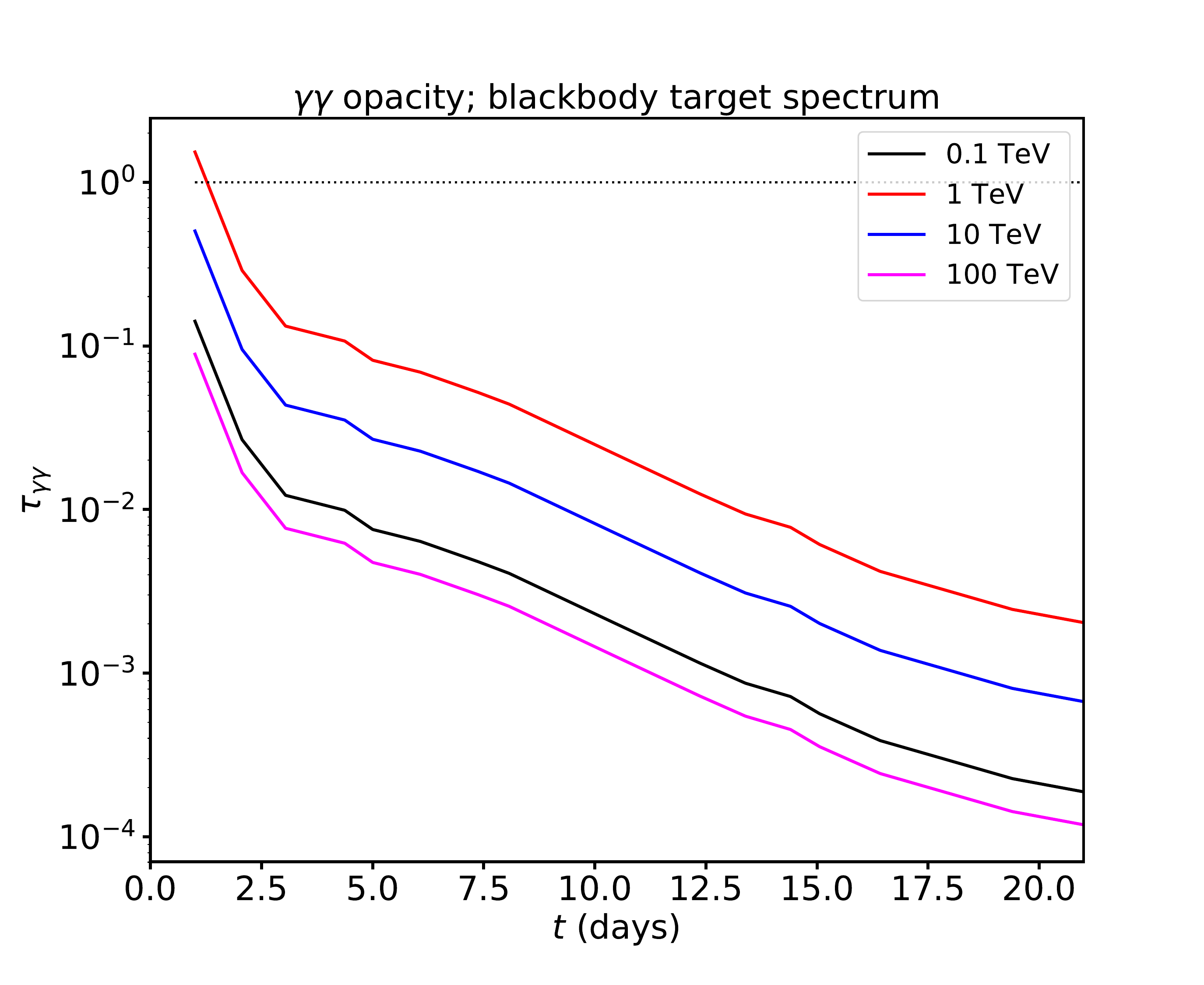}
    \includegraphics[width=8.5cm]{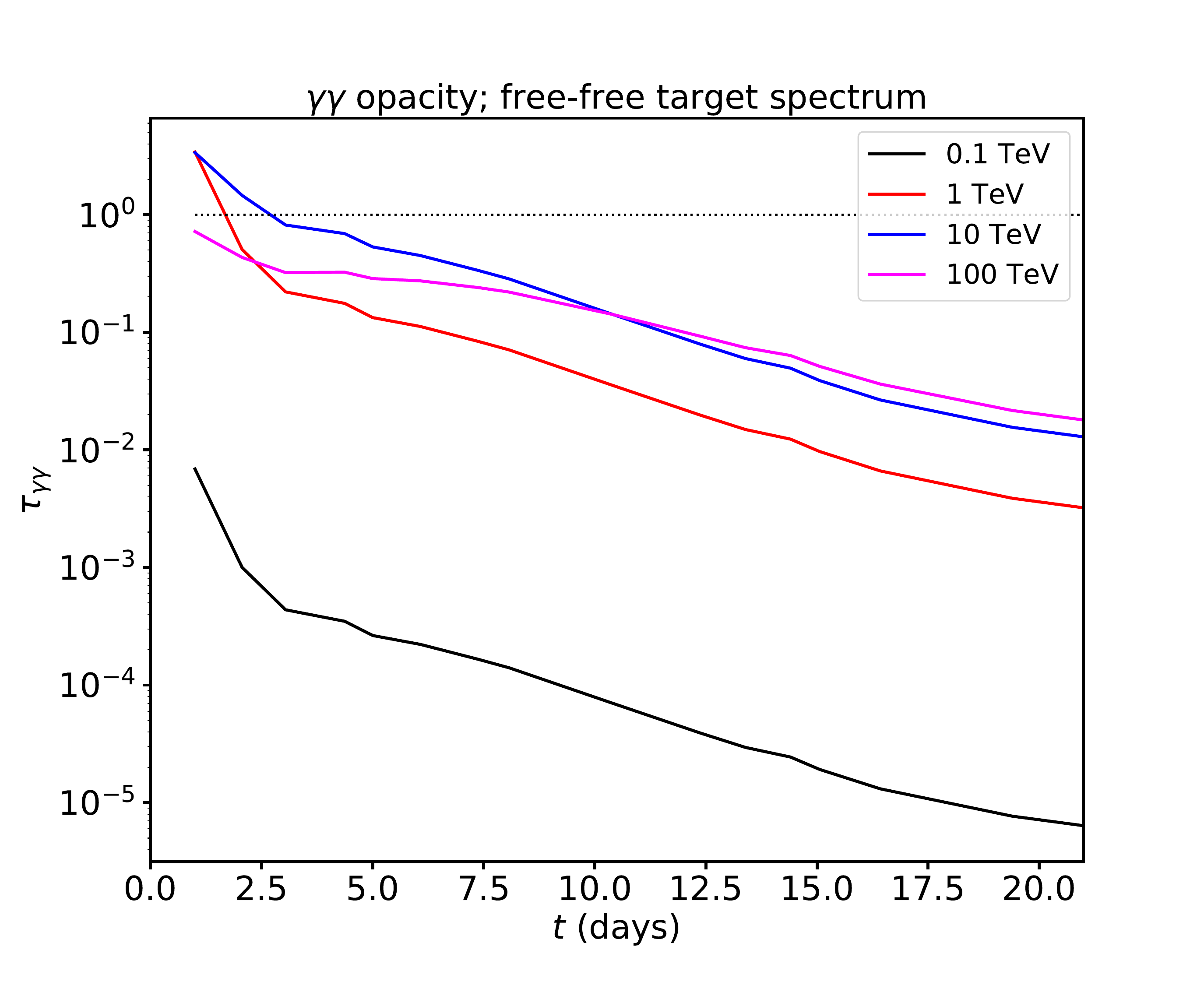}
    \caption{Optical depth in the vicinity of the $\gamma$ ray generating shock $\tau_{\gamma \gamma}$ as a function of time, with different $\gamma$-ray energies shown as lines of different colors as marked. The opacity is due to $\gamma \gamma$ pair creation on the target background radiation of the nova optical light. The results shown in the top panel assume the spectrum of the optical radiation is that of a blackbody at $T_{\rm eff} \approx 8000$ K, while the bottom panel assumes a free-free emission spectrum of gas at the same temperature (accounting for self-absorption at low frequencies).  These two choices roughly bracket the expected level of attenuation for the more realistic but complex optical spectral shape in novae.  Epochs when $\tau_{\gamma\gamma} > 1$ may have their gamma-ray emission strongly attenuated close to the source, where $\tau_{\gamma\gamma} = 1$ is given by the dotted line. }
    \label{fig:tau}
\end{figure}

Figure \ref{fig:tau} shows the 
optical depth, $\tau_{\gamma\gamma}$, as a function of time since the nova eruption, for a photon leaving the vicinity of the shock at a radius $R_{\rm cs} = v_{\rm cs}t$, where $v_{\rm cs} \approx 3800$ km s$^{-1}$ is the intermediate-component velocity estimated from the optical spectrum, thought to trace the shock's cold central shell (see Section \ref{sec:optical}) and hence the location of forward and reverse shocks.  In calculating the value of $\tau_{\gamma\gamma}$, we have made use of the energy density of the optical/near-infrared radiation field,
\begin{equation}
u_{\gamma} = \frac{L_{\rm bol}}{4\pi R_{\rm cs}^{2}c},
\label{eq:ugam}
\end{equation}
estimated from V392 Per’s bolometric light curve $L_{\rm bol}(t)$ (Figure \ref{Fig:bol}).  We separately consider the cases of the optical/infrared spectral energy distribution having the form of a blackbody at temperature $T_{\rm eff} \approx$ 8000 K (top panel of Figure 14) and that of free-free (bremsstrahlung) emission also at temperature $T_{\rm eff} \approx 8000$ K (bottom panel of Figure \ref{fig:tau}). For the effective temperature we avoid using the optical colors to derive a blackbody temperature, given that these colors are heavily affected by the emission lines evolution, and would give an overestimate of the relevant temperature. These two cases (blackbody and free-free) roughly bracket the physically expected range of optical spectral shapes, given the lack of available near-infrared observations of V392 Per to provide additional guidance. For example, \citet{Kato&Hachisu05} argue that the nova emission can be dominated by blackbody emission at early times (during the so-called “fireball” phase) and to later transition to being dominated by free-free emission from a wind or expanding ejecta shell.

The 
optical depth for TeV photons is computed as
\begin{equation}
\tau_{\gamma\gamma}(x) = R_{\rm cs} \int_{1/x} \sigma_{\gamma\gamma}(x,y) \, \frac{dN_{\rm ph}}{dy dV} \, dy,
\end{equation}
where $x = h\nu/m_{\rm e} c ^2$ and $y = h\nu_{\rm opt}/m_{\rm e} c ^2$ are the dimensionless energies of the high-energy and optical (target) photons, respectively, and $\sigma_{\gamma\gamma}$ is the angle-averaged pair production cross-section (e.g., \citealt{Zdziarski_1988}). The target photon spectrum is normalized to the total radiation energy density given by Equation (\ref{eq:ugam}),
\begin{equation}
u_{\gamma} = \int \frac{dN_{\rm ph}}{dy dV} \, m_{\rm e} c^2 y \, dy.
\end{equation}
The shape of the target spectrum follows
$dN_{\rm ph}/(dy dV) \propto y^2/[\exp(m_{\rm e} c^2 y/kT_{\rm eff})-1]$ for a blackbody (Figure \ref{fig:tau}, upper panel) or $dN_{\rm ph}/(dy dV) \propto y^{-1} \exp(-m_{\rm e} c^2 y/kT_{\rm eff})$ for an optically thin bremsstrahlung spectrum (Figure \ref{fig:tau}, lower panel).
It is worth noting that at identical energy densities, the blackbody spectrum places a smaller fraction of target photons at energies $h\nu_{\rm opt} < kT_{\rm eff}$ compared with other plausible physically motivated spectra. As a result, the opacity for photons at $h\nu > (m_{\rm e} c^2)^2/kT_{\rm eff} \sim 1$~TeV is comparatively lower in the blackbody case, as those photons preferentially pair produce on the low-energy tail of the target spectrum.
Note also that at $h\nu \gg (m_{\rm e} c^2)^2/kT_{\rm eff}$, 
the $\gamma$ $\gamma$ opacity behaves approximately as $\tau_{\gamma\gamma} \propto T_{\rm eff}^{-2}$ and $\propto T_{\rm eff}^{-1}$ in the blackbody and free-free cases, respectively.

In the most conservative case of the free-free target spectrum, we see that $\tau_{\gamma\gamma}$ remains larger than unity for a few days after eruption at energies $\gtrsim 1$ TeV. Meanwhile, $\tau_{\gamma\gamma} \lesssim 1$ at all times in the more optimistic case of a blackbody spectrum.  Furthermore, insofar that near the peak of the nova optical light curve the observed emission tends to be dominated by the optically-thick emission from the photosphere instead of optically-thin free-free emission (e.g., from a wind above the photosphere; \citealt{Kato&Hachisu05}), we favor the interpretation that over most, if not all of the time of \Flat detection, V392~Per is transparent to TeV photons.  Still, the \Flat light curve of V392 Per is unusual amongst $\gamma$-ray detected novae for being sharply peaked at early times (Figure \ref{fig:fermi_lc}), and its brightest GeV flux occurs within the first $\sim$2 days of eruption; it is possible that TeV photons were attenuated from V392 Per at these earliest times, when the nova was brightest at GeV energies. In the next section, we consider the shock conditions which would produce very high energy photons in V392 Per.

\subsection{Constraints on the highest energy $\gamma$ rays from nova shocks}\label{sec:emax}

In this section we use 
V392 Per bolometric luminosity and ejecta expansion velocity derived from optical data to estimate the maximum energy to which particles could be accelerated and hence the maximum $\gamma$-ray energy. 

The $\gamma$-ray emission from novae is understood as non-thermal emission from relativistic particles accelerated at shocks (e.g., \citealt{Martin&Dubus13,Ackermann+14}), through the process of diffusive shock acceleration (e.g., \citealt{Blandford&Ostriker78}).  A variety of evidence, from across the electromagnetic spectrum, suggests that the shocks in classical novae are internal to the nova ejecta (e.g., \citealt{Aydi+20,Chomiuk+14,Chomiuk+21araa}), as a fast outflow impacts a slower outflow released earlier in the nova.  On the other hand, in symbiotic novae where the companion is a giant star with dense wind, the shocks may occur as the nova ejecta collides with the external wind (e.g., \citealt{Abdo_etal10}).  V392 Per has an orbital period intermediate between those of cataclysmic variables and symbiotic novae with an atypical radio light curve \citep{Munari+20, Chomiuk+21rad} and hence the nature of the shock interaction$-$internal or external$-$is ambiguous.  However, our discussion to follow regarding the particle acceleration properties is relatively unaffected by this distinction.  

Physical models for the $\gamma$-ray emission divide into ``hadronic'' and ``leptonic'' scenarios depending on whether the emitting particles are primarily relativistic ions or electrons.  Several independent lines of evidence support the hadronic scenario \citep{Chomiuk+21araa}, including: a) the presence of a feature in the $\gamma$-ray spectrum near the pion rest-mass at 135 MeV (e.g., \citealt{Li+17}); b) the non-detection of non-thermal hard X-ray emission by \emph{NuSTAR} (which should be more prominent in leptonic scenarios; 
{\citealt{Vurm&Metzger18,Nelson+19,
Aydi_etal_2020b})}; and c) efficiency limitations on leptonic scenarios due to synchrotron cooling of electrons behind the shock \citep{Li+17}.  Motivated thus, we focus on hadronic scenarios for the $\gamma$ rays.

In the hadronic scenario, relativistic ions collide with ambient ions such as protons, producing pions that decay into $\gamma $rays:
\begin{eqnarray}
p p &\rightarrow& \pi^{0} \rightarrow \gamma \gamma \nonumber \\
&\rightarrow& \pi^{\pm} \rightarrow \mu^{\pm}  \nu_\mu(\bar{\nu}_{\mu}) \rightarrow e^{\pm} +  \nu_e(\bar{\nu}_e)  + \nu_\mu  (\bar{\nu}_\mu).
\end{eqnarray}
Here, $\approx 1/3$ and $\approx 2/3$ of the inelastic p-p collisions go through the $\pi^{0}$ and $\pi^{\pm}$ channels, respectively \citep{Kelner&Aharonian08}. The $\pi^{0}$ channel is expected to dominate the $\gamma$-ray luminosity (e.g., \citealt{Li+17}), but secondary leptons produced via $\pi^{\pm}$ decay can also produce $\gamma$ rays through bremsstrahlung and Inverse Compton processes. A useful rule of thumb is that it requires a proton of energy $10E$ to generate a $\gamma$ ray of energy $E$. Therefore, to produce emission up to the HAWC sensitivity range ($\sim 1-100$ TeV) thus requires 
proton acceleration up to $E_{\rm max} \gtrsim 10-1000$ TeV.  Can nova shocks accelerate particles up to such high energies? 

We consider the shock generated as a fast wind shown of velocity $v_{\rm f} \approx v_3 \approx 5500$ km s$^{-1}$ (see Figure~\ref{Fig:line_profiles}) collides with a slower outflow of velocity $v_{\rm s} \approx v_1 \approx 2700$ km s$^{-1}$, generating an internal shocked shell of velocity $v_{\rm cs} \simeq \xi v_{\rm 1}$ (where the dimensionless parameter $\xi \lesssim 2$, typically; if $v_{\rm cs} = v_{\rm 2} = 3800$ km s$^{-1}$, then $\xi = 1.4$).  Recent studies have shown that the values of $v_{\rm f}$, $v_{\rm s}$, and even $v_{\rm cs}$ may be observed directly in the optical spectra of novae 
{\citep{Aydi_etal_2020b}}, and we have taken our fiducial values here to match those inferred from V392 Per's optical spectra (Section \ref{sec:spec}).


Insofar as an order-unity fraction of the optical nova light is reprocessed thermal emission by the (radiative) reverse shock (e.g., \citealt{Li+17,Aydi+20}), the nova luminosity is related to the mass-loss rate according to
\begin{widetext}
\begin{equation}
L_{\rm bol}(t) \sim L_{\rm sh}(t) \simeq \frac{1}{2}\dot{M}_{\rm f}v_{\rm f}^{2} \nonumber \\
\approx 8\times 10^{36}{\rm erg\,s^{-1}} \frac{\dot{M}_{\rm f}}{10^{-6}M_{\odot}\,\rm yr^{-1}}\left(\frac{v_{\rm f}}{5500\,{\rm km\,s^{-1}}}\right)^{2},
\label{eq:Lsh}
\end{equation}
\end{widetext}
where we have assumed $v_{\rm f} \gg v_{\rm s}$ and treated the fast outflow as a wind of mass-loss rate $\dot{M}_{\rm f}$.

In diffusive shock acceleration, as cosmic rays gain greater and greater energy $E$, they can diffuse back to the shock from a great upstream distance, $z$, because of their larger gyro-radii $r_{\rm g} = E/e B_{\rm sh}$, where 
\begin{widetext}
\begin{equation}
B_{\rm sh} \approx \left(6\pi \epsilon_B m_p n_{\rm f}v_{\rm cs}^{2}\right)^{1/2} \simeq\left(\frac{3}{2}\frac{\epsilon_B \dot{M}_{\rm f}}{v_{\rm f} t^{2}}\right)^{1/2}   \\
 \approx   0.07\,{\rm G}\ \epsilon_{B,-2}^{1/2} \left(\frac{\dot{M}_{\rm f}}{10^{-6}M_{\odot}\,\rm yr^{-1}}\right)^{1/2}\left(\frac{v_{\rm f}}{5500\,\rm km\,s^{-1}}\right)^{-1/2}\left(\frac{t}{1\,{\rm wk}}\right)^{-1},
\end{equation}
\end{widetext}
is the magnetic field behind the reverse shock, for an assumed efficiency of magnetic field amplification $\epsilon_B = \epsilon_{B,-2}\times 10^{-2}$.  This is commensurate with the required field amplification to accelerate ions with an efficiency $\sim 1\%$ \citep{Caprioli&Spitkovsky14b}, as inferred through application of the calorimetric technique \citep{Metzger+15} to correlated $\gamma$-ray/optical emission in novae \citep{Li+17,Aydi+20}.  In the above, we have taken $n_{\rm f} = \dot{M}_{\rm f}/(4\pi m_p R_{\rm cs}^{2}v_{\rm f})$ for the density of the fast outflow at radius $R_{\rm cs} = v_{\rm cs}t$.

The maximum energy to which particles are accelerated before escaping from the vicinity of the shock, $E_{\rm max}$, is found by equating the upstream diffusion time $t_{\rm diff} \sim D/v_{\rm cs}^{2}$, to the minimum of various particle loss timescales.  These include the downstream advection time $t_{\rm adv} \sim z_{\rm acc}/v_{\rm cs}$, where $z_{\rm acc}$ is the width of the acceleration zone, and (in hadronic scenarios) the pion creation timescale $t_{\pi} = (n_{\rm f} \sigma_\pi c)^{-1}$, where $\sigma_{\pi} \sim 2\times 10^{-26}$ cm$^{2}$ is the inelastic cross section for p-p interactions \citep{Kamae+06}.  We consider these limiting processes in turn.

Equating $t_{\rm diff} = t_{\rm adv}$, and taking $D \approx r_{\rm g}c/3$ as the diffusion coefficient \citep{Caprioli&Spitkovsky14b}, one obtains (e.g., \citealt{Metzger+16,Fang+20}) 
\begin{widetext}
\begin{eqnarray}
E_{\rm max} &\sim& \frac{3e B_{\rm sh}v_{\rm cs}z_{\rm acc}}{c}\nonumber \\
E_{\rm max} &\approx& 340\,{\rm TeV}\left(\frac{{z}_{\rm acc}}{R_{\rm cs}}\right)\left(\frac{\xi}{2}\right)^{2}\epsilon_{B,-2}^{1/2}\left(\frac{\dot{M}_{\rm f}}{10^{-6}M_{\odot}\,\rm yr^{-1}}\right)^{1/2} 
\left(\frac{v_{\rm f}}{5500\,{\rm km\,s^{-1}}}\right)^{-1/2}\left(\frac{v_{\rm s}}{2700\,{\rm km\,s^{-1}}}\right)^{2} 
 \label{eq:Emax1}
\end{eqnarray}
where $R_{\rm cs} = v_{\rm cs} t$ is the radius of the shock.

On the other hand, equating $t_{\rm diff} = t_{\pi}$, we obtain
\begin{eqnarray}
E_{\rm max} &\sim& \frac{3e}{c^{2}}\frac{ v_{\rm cs}^{2}B_{\rm sh}}{n_{\rm f}\sigma_{\pi}} \approx 12\pi \frac{m_p}{\sigma_{\pi}}\frac{e}{c^{2}}\frac{B_{\rm sh} v_{\rm f}v_{\rm cs}^{4}t^{2}}{\dot{M}_{\rm f}} \nonumber \\
E_{\rm max} &\approx& 2\times 10^{7}\,{\rm TeV}\,\left(\frac{\xi}{2}\right)^{4}\epsilon_{B,-2}^{1/2}\left(\frac{\dot{M}_{\rm f}}{10^{-6}M_{\odot}\,\rm yr^{-1}}\right)^{-1/2}
\left(\frac{v_{\rm f}}{5500\,{\rm km\,s^{-1}}}\right)^{1/2}\left(\frac{v_{\rm s}}{2700\,{\rm km\,s^{-1}}}\right)^{4}\left(\frac{t}{1\rm\, wk}\right)^{2}
 \label{eq:Emax2}
\end{eqnarray}
\end{widetext}
The maximum energy is given by the minimum of Eqs.~(\ref{eq:Emax1}) and (\ref{eq:Emax2}), which for the system parameters of interest works out to be the former.  In particular, 
taking our fiducial velocity values 
and $L_{\rm opt} \sim 10^{37}-10^{38}$ erg s$^{-1}$ on a timescale of days to weeks (Section \ref{sec:optical}), we see that $\dot{M}_{\rm f} \lesssim 10^{-5}-10^{-6}M_{\odot}$ yr$^{-1}$ (Eq.~\ref{eq:Lsh}).  Thus, from Eq.~(\ref{eq:Emax1}) we infer $E_{\rm max} \lesssim 200-600$ TeV ($z_{\rm acc}/R_{\rm cs}$), in which case we could have expected $\gamma$-ray energies up to $E_{\gamma,\rm max} \sim 0.1E_{\rm max} \sim 20-60$ TeV ($z_{\rm acc}/R_{\rm cs}$).  

If acceleration occurs across a radial scale of order the shock radius (i.e., $z_{\rm acc} \sim R_{\rm cs}$), our estimated $E_{\gamma,\rm max} \sim 20-60$ TeV would appear inconsistent with our constraints on an extension of the measured \Flat spectrum to energies $\gtrsim$ 10 TeV (Section \ref{hypoth}).  
However, various physical effects may reduce the effective extent of the accelerating layer to a width $z_{\rm acc} \ll R_{\rm cs}$ (and hence reduce $E_{\gamma, \rm max}$), such as ion-neutral damping of the \citet{Bell04} instability \citep{Reville+07,Metzger+16} or hydrodynamical thin-shell instabilities of radiative shocks (which corrugate the shock front and alter the effective portion of its surface with the correct orientation relative to the upstream magnetic field to accelerate ions; \citealt{Steinberg&Metzger18}).  The maximum $\gamma$-ray energy generated by the shock could also be lower if the magnetic field amplification factor is less than the fiducial value $\epsilon_{\rm B} = 0.01.$


\section{Conclusions}

The only $\gamma$-ray detected nova in the HAWC data set used in this study is the 2018 eruption of V392 Per.
We present an analysis of the \Flat observations of its GeV $\gamma$-ray signal in Section \ref{sec:fermi}. The \Flat luminosity and spectral shape of V392~Per are typical compared to other \emph{Fermi}-detected novae, but the duration of the $\gamma$ rays was relatively short. Given this, in \S \ref{sec:GeVprops} we revisited the claimed anti-correlation between $\gamma$-ray duration and total emitted energy in the \Flat band \citep{Cheung+16}, and found no such anti-correlation with an improved, larger sample of 15 novae. We do present evidence for a correlation between the duration of the \Flat signal and the optical decline time $t_2$.

HAWC did not detect significant TeV flux in the direction of V392 Per. Therefore, we calculated 95\% confidence upper flux limits for this event, and 
our hypothesis tests disfavor 
{(at 2.8$\sigma$ significance; see Table \ref{tabs:hyptest}) an extension of the \Flat SED to photon energy as high as 10 TeV, and more strongly reject extension to higher energies}. We compared our observations with previous IACT nova studies, and while HAWC is less sensitive, its time agility provides limits during the first week of the GeV emission.

Optical spectroscopy of V392 Per's eruption provides evidence of shocks internal to the nova ejecta, likely occurring between a fast flow expanding at 5500 km s$^{-1}$ and a slow flow of 2700 km s$^{-1}$ (Section \ref{sec:optical})---although we can not rule out the possibility of external shocks with pre-existing circumstellar material. Simple models imply that V392 Per's shocks can accelerate hadrons up to $\sim$400 TeV, potentially yielding $\gamma$ rays of energies up to $\sim$40 TeV (
details depend on complexities like ion-neutral damping; see Section \ref{sec:emax}). 
In Section \ref{sec:tau}, we assess whether very high energy $\gamma$ rays will be observable, given that TeV photons 
are attenuated by pair production on the optical/IR background
at early times. 
For plausible parameters, the nova is expected to be transparent to TeV photons over most of the \Flat detection time window. The non-detection of TeV photons with HAWC is likely attributable to a combination of attenuation at the earliest times (i.e., in the $\sim$first day of eruption, when the GeV $\gamma$ rays are brightest) and the details of diffusive shock acceleration and magnetic field amplification within nova shocks.

HAWC analysis software is undergoing an upgrade which promises both better sensitivity at low energy, and increased field of view.  We will apply the new analysis to V392 Per, RS Oph and several other novae in a future publication.

\acknowledgements
We acknowledge the support from: the US National Science Foundation (NSF); the US Department of Energy Office of High-Energy Physics; the Laboratory Directed Research and Development (LDRD) program of Los Alamos National Laboratory; Consejo Nacional de Ciencia y Tecnolog\'ia (CONACyT), M\'exico, grants 271051, 232656, 260378, 179588, 254964, 258865, 243290, 132197, A1-S-46288, A1-S-22784, c\'atedras 873, 1563, 341, 323, Red HAWC, M\'exico; DGAPA-UNAM grants IG101320, IN111716-3, IN111419, IA102019, IN110621, IN110521; VIEP-BUAP; PIFI 2012, 2013, PROFOCIE 2014, 2015; the University of Wisconsin Alumni Research Foundation; the Institute of Geophysics, Planetary Physics, and Signatures at Los Alamos National Laboratory; Polish Science Centre grant, DEC-2017/27/B/ST9/02272; Coordinaci\'on de la Investigaci\'on Cient\'ifica de la Universidad Michoacana; Royal Society - Newton Advanced Fellowship 180385; Generalitat Valenciana, grant CIDEGENT/2018/034; Chulalongkorn University’s CUniverse (CUAASC) grant; Coordinaci\'on General Acad\'emica e Innovaci\'on (CGAI-UdeG), PRODEP-SEP UDG-CA-499; Institute of Cosmic Ray Research (ICRR), University of Tokyo, H.F. acknowledges support by NASA under award number 80GSFC21M0002. We also acknowledge the significant contributions over many years of Stefan Westerhoff, Gaurang Yodh and Arnulfo Zepeda Dominguez, all deceased members of the HAWC collaboration. Thanks to Scott Delay, Luciano D\'iaz and Eduardo Murrieta for technical support.
CB, LC, and EA are grateful for support from a Cottrell scholarship of the Research Corporation for Science Advancement, NSF grant AST-1751874, and NASA programs Fermi-80NSSC18K1746 and Fermi-80NSSC20K1535.  BDM is supported by the National Science Foundation under grant number 80NSSC20K1561. IV acknowledges support by the ETAg grant PRG1006 and by EU through the ERDF CoE grant TK133. KLL is supported by the Ministry of Science and Technology of the Republic of China (Taiwan) through grant 110-2636-M-006-013, and he is a Yushan (Young) Scholar of the Ministry of Education of the Republic of China (Taiwan).

\bibliography{main.bib}

\begin{thebibliography}{}
\expandafter\ifx\csname natexlab\endcsname\relax\def\natexlab#1{#1}\fi
\providecommand{\url}[1]{\href{#1}{#1}}
\providecommand{\dodoi}[1]{doi:~\href{http://doi.org/#1}{\nolinkurl{#1}}}
\providecommand{\doeprint}[1]{\href{http://ascl.net/#1}{\nolinkurl{http://ascl.net/#1}}}
\providecommand{\doarXiv}[1]{\href{https://arxiv.org/abs/#1}{\nolinkurl{https://arxiv.org/abs/#1}}}

\bibitem[{Abdo {et~al.}(2010)Abdo, Ackermann, Ajello, Atwood, Baldini, Ballet,
  Barbiellini, Bastieri, Bechtol, Bellazzini, Berenji, Blandford, Bloom,
  Bonamente, Borgland, Bouvier, Brandt, Bregeon, Brez, \&
  Yamanaka}]{Abdo_etal10}
Abdo, A., Ackermann, M., Ajello, M., {et~al.} 2010, Science (New York, N.Y.),
  329, 817, \dodoi{10.1126/science.1192537}

\bibitem[{{Abdollahi} {et~al.}(2020){Abdollahi}, {Acero}, {Ackermann},
  {Ajello}, {Atwood}, {Axelsson}, {Baldini}, {Ballet}, {Barbiellini},
  {Bastieri}, {Becerra Gonzalez}, {Bellazzini}, {Berretta}, {Bissaldi}, {Bland
  ford}, {Bloom}, {Bonino}, {Bottacini}, {Brandt}, {Bregeon}, {Bruel},
  {Buehler}, {Burnett}, {Buson}, {Cameron}, {Caputo}, {Caraveo}, {Casandjian},
  {Castro}, {Cavazzuti}, {Charles}, {Chaty}, {Chen}, {Cheung}, {Chiaro},
  {Ciprini}, {Cohen-Tanugi}, {Cominsky}, {Coronado-Bl{\'a}zquez}, {Costantin},
  {Cuoco}, {Cutini}, {D'Ammando}, {DeKlotz}, {Torre Luque}, {de Palma},
  {Desai}, {Digel}, {Lalla}, {Mauro}, {Venere}, {Dom{\'\i}nguez}, {Dumora},
  {Dirirsa}, {Fegan}, {Ferrara}, {Franckowiak}, {Fukazawa}, {Funk}, {Fusco},
  {Gargano}, {Gasparrini}, {Giglietto}, {Giommi}, {Giordano}, {Giroletti},
  {Glanzman}, {Green}, {Grenier}, {Griffin}, {Grondin}, {Grove}, {Guiriec},
  {Harding}, {Hayashi}, {Hays}, {Hewitt}, {Horan}, {J{\'o}hannesson},
  {Johnson}, {Kamae}, {Kerr}, {Kocevski}, {Kovac'evic'}, {Kuss}, {Landriu},
  {Larsson}, {Latronico}, {Lemoine-Goumard}, {Li}, {Liodakis}, {Longo},
  {Loparco}, {Lott}, {Lovellette}, {Lubrano}, {Madejski}, {Maldera},
  {Malyshev}, {Manfreda}, {Marchesini}, {Marcotulli}, {Mart{\'\i}-Devesa},
  {Martin}, {Massaro}, {Mazziotta}, {McEnery}, {Mereu}, {Meyer}, {Michelson},
  {Mirabal}, {Mizuno}, {Monzani}, {Morselli}, {Moskalenko}, {Negro}, {Nuss},
  {Ojha}, {Omodei}, {Orienti}, {Orlando}, {Ormes}, {Palatiello}, {Paliya},
  {Paneque}, {Pei}, {Pe{\~n}a-Herazo}, {Perkins}, {Persic}, {Pesce-Rollins},
  {Petrosian}, {Petrov}, {Piron}, {Poon}, {Porter}, {Principe}, {Rain{\`o}},
  {Rando}, {Razzano}, {Razzaque}, {Reimer}, {Reimer}, {Remy}, {Reposeur},
  {Romani}, {Parkinson}, {Schinzel}, {Serini}, {Sgr{\`o}}, {Siskind}, {Smith},
  {Spandre}, {Spinelli}, {Strong}, {Suson}, {Tajima}, {Takahashi}, {Tak},
  {Thayer}, {Thompson}, {Tibaldo}, {Torres}, {Torresi}, {Valverde}, {Klaveren},
  {Zyl}, {Wood}, {Yassine}, \& {Zaharijas}}]{Abdollahi_etal20}
{Abdollahi}, S., {Acero}, F., {Ackermann}, M., {et~al.} 2020, \apjs, 247, 33,
  \dodoi{10.3847/1538-4365/ab6bcb}

\bibitem[{Abeysekara {et~al.}(2017{\natexlab{a}})Abeysekara, Albert, Alfaro,
  Alvarez, Álvarez, Arceo, Arteaga-Velázquez, Solares, Barber, Baughman, \&
  et~al.}]{Abeysekara_2017a}
Abeysekara, A.~U., Albert, A., Alfaro, R., {et~al.} 2017{\natexlab{a}}, The
  Astrophysical Journal, 843, 40, \dodoi{10.3847/1538-4357/aa7556}

\bibitem[{Abeysekara {et~al.}(2017{\natexlab{b}})Abeysekara, Albert, Alfaro,
  Alvarez, Álvarez, Arceo, Arteaga-Velázquez, Solares, Barber,
  Bautista-Elivar, \& et~al.}]{Abeysekara_2017b}
---. 2017{\natexlab{b}}, The Astrophysical Journal, 843, 39,
  \dodoi{10.3847/1538-4357/aa7555}

\bibitem[{Abeysekara {et~al.}(2019)Abeysekara, Albert, Alfaro, Alvarez,
  {\'{A}}lvarez, Camacho, Arceo, Arteaga-Vel{\'{a}}zquez, Arunbabu, Rojas,
  Solares, Baghmanyan, Belmont-Moreno, BenZvi, Brisbois, Caballero-Mora,
  Capistr{\'{a}}n, Carrami{\~{n}}ana, Casanova, Cotti, Cotzomi, de~Le{\'{o}}n,
  la~Fuente, de~Le{\'{o}}n, Dichiara, Dingus, DuVernois,
  D{\'{\i}}az-V{\'{e}}lez, Ellsworth, Engel, Espinoza, Fick, Fleischhack,
  Fraija, Galv{\'{a}}n-G{\'{a}}mez, Garc{\'{\i}}a-Gonz{\'{a}}lez, Garfias,
  Gonz{\'{a}}lez, Goodman, Harding, Hernandez, Hinton, Hona,
  Hueyotl-Zahuantitla, Hui, Hüntemeyer, Iriarte, Jardin-Blicq, Joshi,
  Kaufmann, Kieda, Lara, Lee, Vargas, Linnemann, Longinotti, Luis-Raya,
  Lundeen, Malone, Marinelli, Martinez, Martinez-Castellanos,
  Mart{\'{\i}}nez-Castro, Mart{\'{\i}}nez-Huerta, Matthews, Miranda-Romagnoli,
  Morales-Soto, Moreno, Mostaf{\'{a}}, Nayerhoda, Nellen, Newbold, Nisa,
  Noriega-Papaqui, Peisker, P{\'{e}}rez-P{\'{e}}rez, Pretz, Ren, Rho,
  Rivi{\`{e}}re, Rosa-Gonz{\'{a}}lez, Rosenberg, Ruiz-Velasco, Salazar, Greus,
  Sandoval, Schneider, Schoorlemmer, Arroyo, Sinnis, Smith, Springer,
  Surajbali, Tabachnick, Tanner, Tibolla, Tollefson, Torres, Weisgarber,
  Westerhoff, Wood, Yapici, Zepeda, \& and}]{Abeysekara_2019}
---. 2019, The Astrophysical Journal, 881, 134,
  \dodoi{10.3847/1538-4357/ab2f7d}

\bibitem[{Abeysekara {et~al.}(2021)}]{brisbois21hal}
Abeysekara, A.~U., {et~al.} 2021, PoS, ICRC2021, 828,
  \dodoi{10.22323/1.395.0828}

\bibitem[{{Ackermann} {et~al.}(2014){Ackermann}, {Ajello}, {Albert}, {Baldini},
  {et~al.}}]{Ackermann+14}
{Ackermann}, M., {Ajello}, M., {Albert}, A., {Baldini}, L., {et~al.} 2014,
  Science, 345, 554, \dodoi{10.1126/science.1253947}

\bibitem[{{Ahnen} {et~al.}(2015){Ahnen}, {Ansoldi}, {Antonelli}, {Antoranz},
  {Babic}, {Banerjee}, {Bangale}, {Barres de Almeida}, {Barrio}, {Becerra
  Gonz{\'a}lez}, {Bednarek}, {Bernardini}, {Biasuzzi}, {Biland}, {Blanch},
  {Bonnefoy}, {Bonnoli}, {Borracci}, {Bretz}, {Carmona}, {Carosi},
  {Chatterjee}, {Clavero}, {Colin}, {Colombo}, {Contreras}, {Cortina},
  {Covino}, {Da Vela}, {Dazzi}, {De Angelis}, {De Caneva}, {De Lotto}, {de
  O{\~n}a Wilhelmi}, {Delgado Mendez}, {Di Pierro}, {Dominis Prester},
  {Dorner}, {Doro}, {Einecke}, {Eisenacher Glawion}, {Elsaesser},
  {Fern{\'a}ndez-Barral}, {Fidalgo}, {Fonseca}, {Font}, {Frantzen}, {Fruck},
  {Galindo}, {Garc{\'\i}a L{\'o}pez}, {Garczarczyk}, {Garrido Terrats}, {Gaug},
  {Giammaria}, {Godinovi{\'c}}, {Gonz{\'a}lez Mu{\~n}oz}, {Guberman},
  {Hanabata}, {Hayashida}, {Herrera}, {Hose}, {Hrupec}, {Hughes}, {Idec},
  {Kellermann}, {Kodani}, {Konno}, {Kubo}, {Kushida}, {La Barbera}, {Lelas},
  {Lewandowska}, {Lindfors}, {Lombardi}, {Longo}, {L{\'o}pez},
  {L{\'o}pez-Coto}, {L{\'o}pez-Oramas}, {Lorenz}, {Majumdar}, {Makariev},
  {Mallot}, {Maneva}, {Manganaro}, {Mannheim}, {Maraschi}, {Marcote},
  {Mariotti}, {Mart{\'\i}nez}, {Mazin}, {Menzel}, {Mirand a}, {Mirzoyan},
  {Moralejo}, {Nakajima}, {Neustroev}, {Niedzwiecki}, {Nievas Rosillo},
  {Nilsson}, {Nishijima}, {Noda}, {Orito}, {Overkemping}, {Paiano}, {Palacio},
  {Palatiello}, {Paneque}, {Paoletti}, {Paredes}, {Paredes-Fortuny}, {Persic},
  {Poutanen}, {Prada Moroni}, {Prandini}, {Puljak}, {Reinthal}, {Rhode},
  {Rib{\'o}}, {Rico}, {Rodriguez Garcia}, {Saito}, {Saito}, {Satalecka},
  {Scapin}, {Schultz}, {Schweizer}, {Sillanp{\"a}{\"a}}, {Sitarek}, {Snidaric},
  {Sobczynska}, {Stamerra}, {Steinbring}, {Strzys}, {Takalo}, {Takami},
  {Tavecchio}, {Temnikov}, {Terzi{\'c}}, {Tescaro}, {Teshima}, {Thaele},
  {Torres}, {Toyama}, {Treves}, {Verguilov}, {Vovk}, {Will}, {Zanin},
  {Desiante}, \& {Hays}}]{Ahnen_etal15}
{Ahnen}, M.~L., {Ansoldi}, S., {Antonelli}, L.~A., {et~al.} 2015, \aap, 582,
  A67, \dodoi{10.1051/0004-6361/201526478}

\bibitem[{Albert {et~al.}(2018)Albert, Alfaro, Alvarez, {\'{A}}lvarez, Arceo,
  Arteaga-Vel{\'{a}}zquez, Rojas, Solares, Bautista-Elivar, Becerril,
  Belmont-Moreno, BenZvi, Bernal, Braun, Brisbois, Caballero-Mora,
  Capistr{\'{a}}n, Carrami{\~{n}}ana, Casanova, Castillo, Cotti, Cotzomi,
  de~Le{\'{o}}n, Le{\'{o}}n, la~Fuente, Hernandez, Dingus, DuVernois,
  D{\'{\i}}az-V{\'{e}}lez, Ellsworth, Engel, Fiorino, Fraija,
  Garc{\'{\i}}a-Gonz{\'{a}}lez, Garfias, Gonz{\'{a}}lez, Goodman, Hampel-Arias,
  Harding, Hernandez, Hernandez-Almada, Hona, Hüntemeyer, Iriarte,
  Jardin-Blicq, Joshi, Kaufmann, Kieda, Lauer, Lennarz, Vargas, Linnemann,
  Longinotti, Proper, Raya, Luna-Garc{\'{\i}}a, L{\'{o}}pez-Coto, Malone,
  Marinelli, Martinez-Castellanos, Mart{\'{\i}}nez-Castro,
  Mart{\'{\i}}nez-Huerta, Matthews, Miranda-Romagnoli, Moreno, Mostaf{\'{a}},
  Nellen, Newbold, Nisa, Noriega-Papaqui, Pelayo, Pretz,
  P{\'{e}}rez-P{\'{e}}rez, Ren, Rho, Rivi{\`{e}}re, Rosa-Gonz{\'{a}}lez,
  Rosenberg, Ruiz-Velasco, Greus, Sandoval, Schneider, Schoorlemmer, Sinnis,
  Smith, Springer, Surajbali, Taboada, Tibolla, Tollefson, Torres, Vianello,
  Weisgarber, Westerhoff, Wood, Yapici, Younk, \& Zhou}]{Albert_2018}
Albert, A., Alfaro, R., Alvarez, C., {et~al.} 2018, The Astrophysical Journal,
  853, 154, \dodoi{10.3847/1538-4357/aaa6d8}

\bibitem[{Albert {et~al.}(2020{\natexlab{a}})Albert, Alfaro, Alvarez,
  Angeles~Camacho, Arteaga-Velázquez, Arunbabu, Avila~Rojas, Ayala~Solares,
  Baghmanyan, Belmont-Moreno, \& et~al.}]{Albert_2020}
---. 2020{\natexlab{a}}, Physical Review Letters, 124,
  \dodoi{10.1103/physrevlett.124.131101}

\bibitem[{Albert {et~al.}(2020{\natexlab{b}})}]{albert_et_al_2020_3hwc}
Albert, A., {et~al.} 2020{\natexlab{b}}, Astrophys. J., 905, 76,
  \dodoi{10.3847/1538-4357/abc2d8}

\bibitem[{Aliu {et~al.}(2012)Aliu, Archambault, Arlen, Aune, Beilicke, Benbow,
  Bouvier, Bradbury, Buckley, Bugaev, \& et~al.}]{Aliu_2012}
Aliu, E., Archambault, S., Arlen, T., {et~al.} 2012, The Astrophysical Journal,
  754, 77, \dodoi{10.1088/0004-637x/754/1/77}

\bibitem[{{Aydi} {et~al.}(2020{\natexlab{a}}){Aydi}, {Sokolovsky}, {Chomiuk},
  {et~al.}}]{Aydi+20}
{Aydi}, E., {Sokolovsky}, K.~V., {Chomiuk}, L., {et~al.} 2020{\natexlab{a}},
  Nature Astronomy, 4, 776, \dodoi{10.1038/s41550-020-1070-y}

\bibitem[{{Aydi} {et~al.}(2020{\natexlab{b}}){Aydi}, {Chomiuk}, {Izzo},
  {Harvey}, {Leahy-McGregor}, {Strader}, {Buckley}, {Sokolovsky}, {Kawash},
  {Kochanek}, {Linford}, {Metzger}, {Mukai}, {Orio}, {Shappee}, {Shishkovsky},
  {Steinberg}, {Swihart}, {Sokoloski}, {Walter}, \& {Woudt}}]{Aydi_etal_2020b}
{Aydi}, E., {Chomiuk}, L., {Izzo}, L., {et~al.} 2020{\natexlab{b}}, \apj, 905,
  62, \dodoi{10.3847/1538-4357/abc3bb}

\bibitem[{{Bell}(2004)}]{Bell04}
{Bell}, A.~R. 2004, \mnras, 353, 550, \dodoi{10.1111/j.1365-2966.2004.08097.x}

\bibitem[{{Blandford} \& {Ostriker}(1978)}]{Blandford&Ostriker78}
{Blandford}, R.~D., \& {Ostriker}, J.~P. 1978, \apjl, 221, L29,
  \dodoi{10.1086/182658}

\bibitem[{{Bode} \& {Evans}(2008)}]{Bode&Evans08}
{Bode}, M.~F., \& {Evans}, A. 2008, {Classical Novae} (Cambridge)

\bibitem[{{Burns} {et~al.}(2011){Burns}, {Stritzinger}, {Phillips}, {Kattner},
  {Persson}, {Madore}, {Freedman}, {Boldt}, {Campillay}, {Contreras},
  {Folatelli}, {Gonzalez}, {Krzeminski}, {Morrell}, {Salgado}, \&
  {Suntzeff}}]{Burns+11}
{Burns}, C.~R., {Stritzinger}, M., {Phillips}, M.~M., {et~al.} 2011, \aj, 141,
  19, \dodoi{10.1088/0004-6256/141/1/19}

\bibitem[{{Caprioli} \& {Spitkovsky}(2014)}]{Caprioli&Spitkovsky14b}
{Caprioli}, D., \& {Spitkovsky}, A. 2014, \apj, 794, 46,
  \dodoi{10.1088/0004-637X/794/1/46}

\bibitem[{{CBAT}(2018)}]{cbat18}
{CBAT}. 2018, CBAT "Transient Object Followup Reports" TCP J04432130+4721280,
  \url{http://www.cbat.eps.harvard.edu/unconf/followups/J04432130+4721280.html}

\bibitem[{{Cheung} {et~al.}(2016){Cheung}, {Jean}, {Shore}, {Stawarz},
  {Corbet}, {Kn{\"o}dlseder}, {Starrfield}, {Wood}, {Desiante}, {Longo},
  {Pivato}, \& {Wood}}]{Cheung+16}
{Cheung}, C.~C., {Jean}, P., {Shore}, S.~N., {et~al.} 2016, \apj, 826, 142,
  \dodoi{10.3847/0004-637X/826/2/142}

\bibitem[{{Chochol} {et~al.}(2021){Chochol}, {Shugarov}, {Hamb{\'a}lek},
  {Skopal}, {Parimucha}, \& {Dubovsk{\'y}}}]{Chochol+21}
{Chochol}, D., {Shugarov}, S., {Hamb{\'a}lek}, {\v{L}}., {et~al.} 2021, in The
  Golden Age of Cataclysmic Variables and Related Objects V, Vol. 2-7, 29.
\newblock \doarXiv{2007.13337}

\bibitem[{{Chodorowski} {et~al.}(1992){Chodorowski}, {Zdziarski}, \&
  {Sikora}}]{Chodorowski+92}
{Chodorowski}, M.~J., {Zdziarski}, A.~A., \& {Sikora}, M. 1992, \apj, 400, 181,
  \dodoi{10.1086/171984}

\bibitem[{{Chomiuk} {et~al.}(2021{\natexlab{a}}){Chomiuk}, {Metzger}, \&
  {Shen}}]{Chomiuk+21araa}
{Chomiuk}, L., {Metzger}, B.~D., \& {Shen}, K.~J. 2021{\natexlab{a}}, \araa,
  59, 391

\bibitem[{{Chomiuk} {et~al.}(2014){Chomiuk}, {Linford}, {Yang}, {O'Brien},
  {Paragi}, {Mioduszewski}, {Beswick}, {Cheung}, {Mukai}, {Nelson}, {Ribeiro},
  {Rupen}, {Sokoloski}, {Weston}, {Zheng}, {Bode}, {Eyres}, {Roy}, \&
  {Taylor}}]{Chomiuk+14}
{Chomiuk}, L., {Linford}, J.~D., {Yang}, J., {et~al.} 2014, \nat, 514, 339,
  \dodoi{10.1038/nature13773}

\bibitem[{{Chomiuk} {et~al.}(2019){Chomiuk}, {Aydi}, {Babul}, {Derdzinski},
  {Kawash}, {Li}, {Linford}, {Metzger}, {Mukai}, {Rupen}, {Sokoloski},
  {Sokolovsky}, \& {Steinberg}}]{Chomiuk_etal19}
{Chomiuk}, L., {Aydi}, E., {Babul}, A.-N., {et~al.} 2019, \baas, 51, 230

\bibitem[{{Chomiuk} {et~al.}(2021{\natexlab{b}}){Chomiuk}, {Linford}, {Aydi},
  {Bannister}, {Krauss}, {Mioduszewski}, {Mukai}, {Nelson}, {Rupen}, {Ryder},
  {Sokoloski}, {Sokolovsky}, {Strader}, {Filipovic}, {Finzell}, {Kawash},
  {Kool}, {Metzger}, {Nyamai}, {Ribeiro}, {Roy}, {Urquhart}, \&
  {Weston}}]{Chomiuk+21rad}
{Chomiuk}, L., {Linford}, J.~D., {Aydi}, E., {et~al.} 2021{\natexlab{b}}, arXiv
  e-prints, arXiv:2107.06251.
\newblock \doarXiv{2107.06251}

\bibitem[{Darnley \& Starrfield(2018)}]{v392progenitor.18}
Darnley, M.~J., \& Starrfield, S. 2018, Res. Notes AAS, 2, 24,
  \dodoi{10.3847/2515-5172/aac26c}

\bibitem[{{Delgado} \& {Hernanz}(2019)}]{Delgado&Hernanz19}
{Delgado}, L., \& {Hernanz}, M. 2019, \mnras, 490, 3691,
  \dodoi{10.1093/mnras/stz2765}

\bibitem[{{Endoh} {et~al.}(2018){Endoh}, {Soma}, {Naito}, \& {Ono}}]{Endoh18}
{Endoh}, I., {Soma}, M., {Naito}, H., \& {Ono}, T. 2018, CBET, 4515

\bibitem[{{Fang} {et~al.}(2020){Fang}, {Metzger}, {Vurm}, {Aydi}, \&
  {Chomiuk}}]{Fang+20}
{Fang}, K., {Metzger}, B.~D., {Vurm}, I., {Aydi}, E., \& {Chomiuk}, L. 2020,
  arXiv e-prints, arXiv:2007.15742.
\newblock \doarXiv{2007.15742}

\bibitem[{{Franckowiak} {et~al.}(2018){Franckowiak}, {Jean}, {Wood}, {Cheung},
  \& {Buson}}]{Franckowiak+18}
{Franckowiak}, A., {Jean}, P., {Wood}, M., {Cheung}, C.~C., \& {Buson}, S.
  2018, \aap, 609, A120, \dodoi{10.1051/0004-6361/201731516}

\bibitem[{{Friedjung}(1987)}]{Friedjung87}
{Friedjung}, M. 1987, \aap, 180, 155

\bibitem[{{Friedman} {et~al.}(2011){Friedman}, {York}, {McCall}, {Dahlstrom},
  {Sonnentrucker}, {Welty}, {Drosback}, {Hobbs}, {Rachford}, \&
  {Snow}}]{Friedman_etal_2011}
{Friedman}, S.~D., {York}, D.~G., {McCall}, B.~J., {et~al.} 2011, \apj, 727,
  33, \dodoi{10.1088/0004-637X/727/1/33}

\bibitem[{{Gaia Collaboration} {et~al.}(2016){Gaia Collaboration}, {Prusti},
  {de Bruijne}, {Brown}, {Vallenari}, {Babusiaux}, {Bailer-Jones}, {Bastian},
  {Biermann}, {Evans}, \& et~al.}]{Gaia16}
{Gaia Collaboration}, {Prusti}, T., {de Bruijne}, J.~H.~J., {et~al.} 2016,
  \aap, 595, A1, \dodoi{10.1051/0004-6361/201629272}

\bibitem[{{Gaia Collaboration} {et~al.}(2021){Gaia Collaboration}, {Brown},
  {Vallenari}, {Prusti}, {de Bruijne}, {Babusiaux}, {Biermann}, {Creevey},
  {Evans}, {Eyer}, \& et~al.}]{Gaia21}
{Gaia Collaboration}, {Brown}, A.~G.~A., {Vallenari}, A., {et~al.} 2021, \aap,
  649, A1, \dodoi{10.1051/0004-6361/202039657}

\bibitem[{{Gallagher} \& {Starrfield}(1976)}]{Gallagher_Starrfield_1976}
{Gallagher}, J.~S., \& {Starrfield}, S. 1976, \mnras, 176, 53,
  \dodoi{10.1093/mnras/176.1.53}

\bibitem[{{Gallagher} \& {Starrfield}(1978)}]{Gallagher&Starrfield78}
---. 1978, \araa, 16, 171, \dodoi{10.1146/annurev.aa.16.090178.001131}

\bibitem[{{Gordon} {et~al.}(2021){Gordon}, {Aydi}, {Page}, {Li}, {Chomiuk},
  {Sokolovsky}, {Mukai}, \& {Seitz}}]{Gordon+21}
{Gordon}, A.~C., {Aydi}, E., {Page}, K.~L., {et~al.} 2021, \apj, 910, 134,
  \dodoi{10.3847/1538-4357/abe547}

\bibitem[{{Hachisu} \& {Kato}(2004)}]{Hachisu_Kato_2004}
{Hachisu}, I., \& {Kato}, M. 2004, \apjl, 612, L57, \dodoi{10.1086/424595}

\bibitem[{{Kafka}(2020)}]{Kafka_2020}
{Kafka}, S. 2020, Observations from the AAVSO International Database, \url{
  https://www.aavso.org}

\bibitem[{{Kamae} {et~al.}(2006){Kamae}, {Karlsson}, {Mizuno}, {Abe}, \&
  {Koi}}]{Kamae+06}
{Kamae}, T., {Karlsson}, N., {Mizuno}, T., {Abe}, T., \& {Koi}, T. 2006, \apj,
  647, 692, \dodoi{10.1086/505189}

\bibitem[{{Kato} \& {Hachisu}(2005)}]{Kato&Hachisu05}
{Kato}, M., \& {Hachisu}, I. 2005, \apjl, 633, L117, \dodoi{10.1086/498300}

\bibitem[{{Kelner} \& {Aharonian}(2008)}]{Kelner&Aharonian08}
{Kelner}, S.~R., \& {Aharonian}, F.~A. 2008, \prd, 78, 034013,
  \dodoi{10.1103/PhysRevD.78.034013}

\bibitem[{{Li} {et~al.}(2018){Li}, {Chomiuk}, \& {Strader}}]{Li_etal18}
{Li}, K.-L., {Chomiuk}, L., \& {Strader}, J. 2018, The Astronomer's Telegram,
  11590

\bibitem[{{Li} {et~al.}(2017){Li}, {Metzger}, {Chomiuk}, {Vurm}, {Strader},
  {Finzell}, {Beloborodov}, {Nelson}, {Shappee}, {Kochanek}, {Prieto}, {Kafka},
  {Holoien}, {Thompson}, {Luckas}, \& {Itoh}}]{Li+17}
{Li}, K.-L., {Metzger}, B.~D., {Chomiuk}, L., {et~al.} 2017, Nature Astronomy,
  1, 697, \dodoi{10.1038/s41550-017-0222-1}

\bibitem[{{Martin} \& {Dubus}(2013)}]{Martin&Dubus13}
{Martin}, P., \& {Dubus}, G. 2013, \aap, 551, A37,
  \dodoi{10.1051/0004-6361/201220289}

\bibitem[{{Martin} {et~al.}(2018){Martin}, {Dubus}, {Jean}, {Tatischeff}, \&
  {Dosne}}]{Martin+18}
{Martin}, P., {Dubus}, G., {Jean}, P., {Tatischeff}, V., \& {Dosne}, C. 2018,
  \aap, 612, A38, \dodoi{10.1051/0004-6361/201731692}

\bibitem[{{Mathis}(1990)}]{Mathis90}
{Mathis}, J.~S. 1990, \araa, 28, 37,
  \dodoi{10.1146/annurev.aa.28.090190.000345}

\bibitem[{{McLaughlin}(1943)}]{McLaughlin43}
{McLaughlin}, D.~B. 1943, Publications of Michigan Observatory, 8, 149

\bibitem[{{Mclaughlin}(1947)}]{Mclaughlin47}
{Mclaughlin}, D.~B. 1947, \pasp, 59, 244, \dodoi{10.1086/125957}

\bibitem[{{Metzger} {et~al.}(2016){Metzger}, {Caprioli}, {Vurm}, {Beloborodov},
  {Bartos}, \& {Vlasov}}]{Metzger+16}
{Metzger}, B.~D., {Caprioli}, D., {Vurm}, I., {et~al.} 2016, \mnras, 457, 1786,
  \dodoi{10.1093/mnras/stw123}

\bibitem[{{Metzger} {et~al.}(2015){Metzger}, {Finzell}, {Vurm}, {Hasco{\"e}t},
  {Beloborodov}, \& {Chomiuk}}]{Metzger+15}
{Metzger}, B.~D., {Finzell}, T., {Vurm}, I., {et~al.} 2015, \mnras, 450, 2739,
  \dodoi{10.1093/mnras/stv742}

\bibitem[{{Metzger} {et~al.}(2014){Metzger}, {Hasco{\"e}t}, {Vurm},
  {Beloborodov}, {Chomiuk}, {Sokoloski}, \& {Nelson}}]{Metzger+14}
{Metzger}, B.~D., {Hasco{\"e}t}, R., {Vurm}, I., {et~al.} 2014, \mnras, 442,
  713, \dodoi{10.1093/mnras/stu844}

\bibitem[{{Munari} {et~al.}(2020){Munari}, {Moretti}, \& {Maitan}}]{Munari+20}
{Munari}, U., {Moretti}, S., \& {Maitan}, A. 2020, \aap, 639, L10,
  \dodoi{10.1051/0004-6361/202038403}

\bibitem[{{Nelson} {et~al.}(2019){Nelson}, {Mukai}, {Li}, {Vurm}, {Metzger},
  {Chomiuk}, {Sokoloski}, {Linford}, {Bohlsen}, \& {Luckas}}]{Nelson+19}
{Nelson}, T., {Mukai}, K., {Li}, K.-L., {et~al.} 2019, \apj, 872, 86,
  \dodoi{10.3847/1538-4357/aafb6d}

\bibitem[{{Poznanski} {et~al.}(2012){Poznanski}, {Prochaska}, \&
  {Bloom}}]{Poznanski_etal_2012}
{Poznanski}, D., {Prochaska}, J.~X., \& {Bloom}, J.~S. 2012, \mnras, 426, 1465,
  \dodoi{10.1111/j.1365-2966.2012.21796.x}

\bibitem[{{Reville} {et~al.}(2007){Reville}, {Kirk}, {Duffy}, \&
  {O'Sullivan}}]{Reville+07}
{Reville}, B., {Kirk}, J.~G., {Duffy}, P., \& {O'Sullivan}, S. 2007, \aap, 475,
  435, \dodoi{10.1051/0004-6361:20078336}

\bibitem[{{Schaefer}(2018)}]{Schaefer_2018}
{Schaefer}, B.~E. 2018, \mnras, 481, 3033, \dodoi{10.1093/mnras/sty2388}

\bibitem[{Schaefer(2021)}]{v392.period}
Schaefer, B.~E. 2021, Res. Notes AAS, 5, 150.
\newblock \doarXiv{2106.13907}

\bibitem[{Smith(2016)}]{Smith_2015}
Smith, A.~J. 2016, PoS, ICRC2015, 966, \dodoi{10.22323/1.236.0966}

\bibitem[{{Steinberg} \& {Metzger}(2018)}]{Steinberg&Metzger18}
{Steinberg}, E., \& {Metzger}, B.~D. 2018, \mnras, 479, 687,
  \dodoi{10.1093/mnras/sty1641}

\bibitem[{Surajbali(2021)}]{pooja}
Surajbali, P. 2021, PhD thesis, University of Heidelberg

\bibitem[{{Teyssier}(2019)}]{Teyssier_2019}
{Teyssier}, F. 2019, Contributions of the Astronomical Observatory Skalnate
  Pleso, 49, 217

\bibitem[{Vianello {et~al.}(2016)Vianello, Lauer, Younk, Tibaldo, Burgess,
  Ayala~Solares, Harding, Hui, Omodei, \& Zhou}]{vianello:3ml.2015}
Vianello, G., Lauer, R., Younk, P., {et~al.} 2016, PoS, ICRC2015, 1042,
  \dodoi{10.22323/1.236.1042}

\bibitem[{{Vurm} \& {Metzger}(2018)}]{Vurm&Metzger18}
{Vurm}, I., \& {Metzger}, B.~D. 2018, \apj, 852, 62,
  \dodoi{10.3847/1538-4357/aa9c4a}

\bibitem[{{Wagner} {et~al.}(2018{\natexlab{a}}){Wagner}, {Terndrup}, {Darnley},
  {Starrfield}, {Woodward}, \& {Henze}}]{Wagneretal_2018}
{Wagner}, R.~M., {Terndrup}, D., {Darnley}, M.~J., {et~al.} 2018{\natexlab{a}},
  The Astronomer's Telegram, 11588, 1

\bibitem[{{Wagner} {et~al.}(2018{\natexlab{b}}){Wagner}, {Terndrup}, {Darnley},
  {Starrfield}, {Woodward}, \& {Henze}}]{Wagner+18}
---. 2018{\natexlab{b}}, The Astronomer's Telegram, 11588, 1

\bibitem[{Younk {et~al.}(2016)Younk, Lauer, Vianello, Harding, Ayala~Solares,
  Zhou, \& Hui}]{younk2015highlevel}
Younk, P.~W., Lauer, R.~J., Vianello, G., {et~al.} 2016, PoS, ICRC2015, 948,
  \dodoi{10.22323/1.236.0948}

\bibitem[{{Zdziarski}(1988)}]{Zdziarski_1988}
{Zdziarski}, A.~A. 1988, \apj, 335, 786, \dodoi{10.1086/166967}

\end{thebibliography}
\end{document}